\documentclass{aa}  
\usepackage{graphicx}
\graphicspath{{Figures/}}
\usepackage{txfonts}
\usepackage{hyperref}
\usepackage{pgf}   
\usepackage{siunitx}
\usepackage{makecell}

\hypersetup{
  colorlinks   = true, 
  urlcolor     = blue, 
  linkcolor    = blue, 
  citecolor   = blue
}

\newcommand{\Swift}{\textit{Swift}}
\newcommand{\Fermi}{\textit{Fermi}}

\begin{document} 

   \title{The triple-peaked afterglow of GRB 210731A from X-ray to radio frequencies\thanks{Table \ref{tab:photometry} is available in electronic form at the CDS via anonymous ftp to cdsarc.cds.unistra.fr (130.79.128.5) or via https://cdsarc.cds.unistra.fr/cgi-bin/qcat?J/A+A}}

   \author{S. de Wet \inst{1,}\thanks{Corresponding author \email{DWTSIM002@myuct.ac.za}}
          \and T. Laskar\inst{2,3} \and P.J. Groot \inst{1,3,4} \and F. Cavallaro\inst{1,5} \and A. Nicuesa Guelbenzu\inst{6}
         \and S. Chastain\inst{7} \and L. Izzo\inst{8} \and A. Levan\inst{3} \and D.B. Malesani\inst{3,9,10} \and I.M. Monageng\inst{1,4} \and A.J. van der Horst\inst{7} \and W. Zheng\inst{11} \and S. Bloemen\inst{3} \and A.V. Filippenko\inst{11} \and D.A. Kann\inst{12} \and S. Klose\inst{6} \and D.L.A. Pieterse\inst{3} \and A. Rau\inst{13} \and P.M. Vreeswijk\inst{3} \and P. Woudt\inst{1} \and Z.-P. Zhu\inst{14}
          }
          
            \institute{Inter-University Institute for Data Intensive Astronomy \& Department of Astronomy, University of Cape Town, Private Bag X3, Rondebosch, 7701, South Africa
              \and
            Department of Physics \& Astronomy, University of Utah, Salt Lake City, UT 84112, USA
            \and
              Department of Astrophysics/IMAPP, Radboud University, P.O. Box 9010, 6500 GL, Nijmegen, The Netherlands
              \and 
              South African Astronomical Observatory, P.O. Box 9, 7935, Observatory, South Africa
             \and
             INAF-Osservatorio Astrofisico di Catania, Via S. Sofia 78, I-95123 Catania, Italy
             \and
             Th\"uringer Landessternwarte Tautenburg, Sternwarte 5, 07778 Tautenburg, Germany
             \and
             Department of Physics, The George Washington University, 725 21\textsuperscript{st} Street NW, Washington, DC 20052, USA
             \and
             DARK, Niels Bohr Institute, University of Copenhagen, Jagtvej 128, 2200 Copenhagen, Denmark
             \and
             Cosmic Dawn Center (DAWN), Denmark
            \and
            Niels Bohr Institute, University of Copenhagen, R{\aa}dmandsgade 62-64, 2200, Copenhagen N, Denmark
             \and
             Department of Astronomy, University of California, Berkeley, CA 94720-3411, USA
             \and
            Hessian Research Cluster ELEMENTS, Giersch Science Center, Max-von-Laue-Strasse 12, Goethe University Frankfurt, Campus Riedberg, 60438 Frankfurt am Main, Germany
             \and
             Max-Planck-Institut f\"ur Extraterrestrische Physik, Giessenbachstra\ss{}e, 85748 Garching, Germany
             \and 
             Department of Astronomy, School of Physics, Huazhong University of Science and Technology, Wuhan, 430074, China
             }

   \date{Received \today; accepted \today}
   
  \abstract
   {GRB 210731A was a long-duration ($T_{90}=22.5$ s) gamma-ray burst discovered by the Burst Alert Telescope (BAT) aboard the \textit{Neil Gehrels} \Swift{} observatory. \Swift{} triggered the wide-field, robotic MeerLICHT optical telescope in Sutherland; it began observing the BAT error circle 286~seconds after the \Swift{} trigger and discovered the optical afterglow of GRB 210731A in its first 60-second $q$-band exposure. Multi-colour observations of the afterglow with MeerLICHT revealed a light curve that showed three peaks of similar brightness within the first four hours. The unusual optical evolution prompted multi-wavelength follow-up observations that spanned from X-ray to radio frequencies.}
   {We present the results of our follow-up campaign and interpret our observations in the framework of the synchrotron forward shock model.}
   {We performed temporal and spectral fits to determine the spectral regime and external medium density profile, and performed detailed multi-wavelength theoretical modelling of the afterglow following the last optical peak at ${\sim}0.2$ days to determine the intrinsic blast wave parameters.}
   {We find a preference for a stellar wind density profile consistent with a massive star origin, while our theoretical modelling results in fairly typical shock microphysics parameters. Based on the energy released in $\gamma$ rays and the kinetic energy in the blast wave, we determine a low radiative efficiency of $\eta\approx0.02$. The first peak in the optical light curve is likely the onset of the afterglow. We find that energy injection into the forward shock offers the simplest explanation for the subsequent light curve evolution, and that the blast wave kinetic energy increasing by a factor of ${\sim}1000$ from the first peak to the last peak is indicative of substantial energy injection. Our highest-likelihood theoretical model over-predicts the 1.4~GHz flux by a factor of approximately three with respect to our upper limits, possibly implying a population of thermal electrons within the shocked region.}
 {}
\maketitle

\section{Introduction}
Gamma-ray bursts (GRBs) are the most energetic explosions in the Universe, with isotropic $\gamma$-ray energies of up to ${\sim}10^{55}$~erg \citep{Kumar2015}. Long-duration bursts are typically associated with the core collapse of massive stars \citep{Colgate1968,Woosley1993,vanparadijs1997,Galama1998a,Woosley2006}, where the compact object remnant acts as a central engine powering a collimated relativistic jet. In the fireball shock model \citep{Rees1992,Meszaros1993,Meszaros1997,Sari1998}, internal shocks within the expanding ejecta are the source of the prompt $\gamma$-ray emission and the interaction of the relativistic outflow with the circumburst medium (the external forward shock) is the source of the afterglow emission.

Prior to the launch of the \textit{Neil Gehrels} \Swift{} Observatory \citep{Swift} in 2004, observations of afterglows showed broad agreement with the basic external shock model \citep{Galama1998b,Panaitescu2001,Panaitescu2002,Yost2003}. Rapid X-ray and UV/optical follow-up of GRB triggers with \Swift{} have revealed rich features in early-time X-ray and optical light curves that challenge the standard theory. Similar to the `canonical' X-ray light curve \citep{Zhang2006,Nousek2006}, optical light curves have been decomposed into a number of distinct components that arise from different emission sites and physical mechanisms \citep[see the synthetic light curve in][]{Li2012}. In addition to the normal and jet break decay segments explained by the standard forward shock model, onset bumps, steep decay segments, flares, and late-time re-brightenings have been observed in optical afterglows. Early onset bumps with a smooth transition to the normal decay segment are regarded as the onset of afterglow and were predicted within the standard theory \citep{Sari1999,Kobayashi2007}; steep decay segments early on (some with an additional steep rise) have been attributed to reverse shock emission \citep{Meszaros1997,Sari1999,Zhang2003,Yi2020}; similar to X-ray light curves, shallow decay and plateau segments and flares have been observed in a number of bursts \citep{Mangano2007,Greiner2009,Swenson2013}; late-time re-brightenings have also been observed in some optical afterglows \citep{Nardini2011,Liang2013}, with a few having rarer simultaneous X-ray re-brightenings \citep[e.g. GRB 120326A;][]{Melandri2014,Urata2014,Hou2014,Laskar2015}. Furthermore, small-scale bumps and wiggles have been seen superposed over the larger-scale light curve features \citep[e.g. GRBs 021004 and 030329;][]{Holland2003,Lipkin2004}. Proposed explanations for these additional features include inhomogeneities in the circumburst medium, multiple-component jets, structured jets, varying microphysical parameters, and energy injection into the forward shock \citep{Lazzati2002,Dai2003,Mundell2007,Racusin2008,Filgas2011,Meszaros1998,Zhang2002a,Rossi2002,Ioka2006,Fan2006,Granot2006,Sari2000,Zhang2002b,Bjornsson2004,Laskar2015,Lin2018}. 

Here we report on multi-wavelength observations of the long-duration gamma-ray burst GRB 210731A discovered by \Swift{} \citep{Swift}. Our early-time MeerLICHT optical observations show a complex light curve evolution with an initial smooth bump followed by two further re-brightenings. We combine our observations with X-ray and radio data that span 200~seconds to 118~days post-trigger and interpret our observations in the traditional synchrotron forward shock framework. We investigate the nature of the early optical light curve evolution. 

We adopt a $\Lambda$ cold dark matter cosmology with $\Omega_m=0.31$, $\Omega_{\Lambda}=0.69$, and $H_0=68$~km~s$^{-1}$~Mpc$^{-1}$ \citep{Planck2016}. All reported magnitudes are in the AB magnitude system unless stated otherwise, and errors are reported at the $1\sigma$ level.  
\section{Observations}\label{sec:data}
\subsection{Prompt gamma-ray emission}\label{sec:GRB}
The \Swift{} Burst Alert Telescope \citep[BAT;][]{BAT} was triggered by GRB 210731A at 22:21:08 UT on 2021 July 31, with the mask-weighted 15--350~keV light curve showing a single-pulse structure of duration $T_{90}=22.5\pm2.8$~s \citep{GCN30580}, making GRB 210731A a long-duration GRB under the traditional $>2$~s duration limit. GRB 210731A triggered the \Fermi{} Gamma-ray Burst Monitor \citep[GBM;][]{Fermi,GCN30573} one second earlier than \Swift{}/BAT, with the 10-1000 keV light curve showing a single pulse with duration\footnote{Obtained from the online \href{https://heasarc.gsfc.nasa.gov/W3Browse/fermi/fermigbrst.html}{Fermi GBM Burst Catalog} \citep{vonkienlin2020}.} $T_{90}=25.9\pm5.3$~s, in agreement with the BAT duration. The time-averaged spectra for both BAT and GBM were best-fitted with a power law function and exponential high-energy cutoff with photon indices of $-0.25\pm0.59$ and $-0.1\pm0.1$, and cutoff energies of $(107\pm27)$~keV and $(175\pm11)$~keV, respectively \citep{GCN30580,GCN30573}. The 10-1000 keV GBM fluence\footnote{See footnote 1.} integrated over the burst duration was $(3.05\pm0.06)\times10^{-6}$~erg~cm$^{-2}$. Using a measured afterglow redshift of $z=1.2525$ obtained by X-Shooter on the Very Large Telescope at 1.19~days post-trigger \citep{GCN30583}, this corresponds to an isotropic-equivalent $\gamma$-ray energy of $E_{\gamma,\mathrm{iso}}=(1.29\pm0.03)\times10^{52}$~erg.  

We take the \Swift{}/BAT trigger time as $T_0$ for this burst and reference all future times with respect to this $T_0$.  

\subsection{X-ray observations}\label{sec:XRT}
The \Swift{} X-Ray Telescope \citep[XRT;][]{XRT} started observing the field of GRB 210731A 201 seconds post-trigger, finding a bright new X-ray source consistent with the BAT position \citep{GCN30568}. The initial 62 seconds of data were obtained in windowed timing (WT) mode after which \Swift{} had to slew away. Data capture recommenced in photon counting (PC) mode at 3.3 hours post-trigger. We obtained the X-ray light curve and spectra from the online Swift-XRT GRB Catalogue\footnote{The Burst Analyser for GRB 210731A is available on the \href{https://www.swift.ac.uk/burst_analyser/01062336/}{UK Swift Data Science Centre website}.} \citep{Evans2007,Evans2009}. The Burst Analyser \citep{Evans2010} count-rate light curve showed that the X-ray flux was decreasing rapidly during the WT-mode observations with a spectrum that hardened from a photon index of $\Gamma_\mathrm{X}=3.2$ to 2.2 over 60 seconds. Once data capture resumed in PC mode at 10~ks, the X-ray light curve was in a shallow decay phase before declining more steeply at ${\sim}20$~ks post-trigger.

We fitted the PC-mode spectrum with a photoelectrically absorbed power-law model (\verb|tbabs*ztbabs*pow|) in Xspec version 12.12.0, fixing the source redshift at 1.2525 and Galactic hydrogen column density at $N_\mathrm{H}^\mathrm{Gal}=1.15\times10^{21}$~cm$^{-2}$ for consistency with the online fit. The fitted spectrum was characterised by a photon index of $\Gamma=2.00_{-0.11}^{+0.11}$ with a host galaxy column density of $N_\mathrm{H}^\mathrm{host}=2.46_{-1.56}^{+1.99}\times10^{21}$~cm$^{-2}$ and C-stat 182.1 for 213 degrees of freedom. There were insufficient photons in the PC-mode light curve for time-resolved analysis and to test for spectral evolution. We converted the PC-mode count-rate light curve to a 1 keV flux density light curve using the spectral index from our PC-mode spectral fit of $\beta_\mathrm{X}\equiv1-\Gamma_\mathrm{X}\approx-1.00$ and the online unabsorbed count-to-flux conversion factor of $4.36\times10^{-11}$~erg~cm$^{-2}$~ct$^{-1}$. We performed a similar procedure for the WT-mode data using the spectral parameters in the automated fit on the \Swift{} website, with a photon index of 3.07 and unabsorbed count-to-flux conversion factor of $4.69\times10^{-11}$~erg~cm$^{-2}$~ct$^{-1}$. 

\subsection{Optical/near-infrared observations}\label{sec:opt}
The fully robotic, 60~cm MeerLICHT optical telescope \citep{Bloemen2016} was automatically triggered by \Swift{}/BAT and began observing the field of GRB 210731A 286~seconds after the BAT trigger, taking 60~second exposures in the $u,g,r,i,z,$ and $q$ optical bands (where the $q$ band is roughly equivalent to $g+r$), following the sequence $quqgqrqiqz$ in order to obtain high cadence coverage in the wider and more sensitive $q$ band with quasi-simultaneous multi-colour coverage of the evolving afterglow. Comparison of the first $q$-band image with an existing MeerLICHT reference image revealed a new transient candidate at $\alpha=20^\mathrm{h}01^\mathrm{m}13.19^\mathrm{s}$, $\delta = -28^\mathrm{d}03^\mathrm{m}40.10^\mathrm{s}$ (J2000). This position was $0.3^{\prime\prime}$ away from the refined XRT position \citep{GCN30569}, confirming the new source as the optical afterglow of GRB 210731A \citep{GCN30570}. These observations continued until the target set, approximately 4.29~hours post-trigger. Four cycles of the same filter sequence were obtained the following night in two time intervals separated by $\sim$2~hours. Since the afterglow was by this point in a declining phase and below the 60~second single-exposure detection limit, repeated $q$-band exposures were taken on the nights of 2021 August 2 and 3 in order to track the optical light curve. The MeerLICHT pipeline (Vreeswijk et al., in prep) was used to perform standard charge-coupled device (CCD) reduction tasks as well as astrometry and point-spread function (PSF) photometry, producing a catalogue file containing all 5$\sigma$ source detections for each image. For images where the afterglow was fainter than 5$\sigma$ above the background we used forced photometry to obtain magnitudes that were at least at the 3$\sigma$ level. Images from the night of 2021 August 1 onwards were co-added to produce more significant detections or deeper upper limits. 

The \Swift{} UltraViolet and Optical Telescope \citep[UVOT;][]{UVOT} took a single 61.7~second exposure in the $white$ filter beginning 210.4~seconds after the BAT trigger but did not continue observing the field of the GRB until 3.27~hours after the trigger, whereafter it was observed with multiple filters intermittently over the next five days \citep{GCN30568,GCN30572}. We performed aperture photometry on the \Swift{}/UVOT data using standard analysis tools from the HEASoft \citep{Heasoft} \Swift{} FTOOLS software package (version 6.29c\footnote{Available at \href{https://heasarc.gsfc.nasa.gov/docs/software/lheasoft/}{https://heasarc.gsfc.nasa.gov/docs/software/lheasoft/}}). We extracted magnitudes using the tool \verb|uvotsource| with a $3.5^{\prime\prime}$ radius aperture centred on the afterglow position, and a nearby background aperture with a $10^{\prime\prime}$ radius. A total of 64 individual exposures in all seven UVOT filters were taken over the course of the follow-up campaign. We co-added exposures in the same filter with clear detections but taken close to each other temporally using \verb|uvotimsum| in order to produce more significant detections, and once the afterglow became too faint to detect in individual exposures we co-added images within wider time baselines in order to provide the deepest limiting magnitudes. 

The afterglow of GRB 210731A was observed simultaneously in the $g'r'i'z'JHK$ bands with the Gamma-Ray Burst Optical Near-Infrared Detector \citep[GROND;][]{GROND} mounted at the 2.2~m MPG telescope at the European Southern Observatory (ESO) La Silla Observatory in Chile. The afterglow was clearly detected in all bands in the first epoch of observations taken 4.2~hours after the GRB trigger \citep{GCN30574}. A further three epochs were obtained at 1.225, 2.214, and 5.253~days post-trigger \citep{GCN30584}. We also obtained deep host-galaxy observations at 285~days that yielded detections in the $g'$ and $r'$ bands, and which we regard as the host-galaxy flux. The multi-colour GROND data were analysed through standard PSF photometry using DAOPHOT \citep{Stetson1987} and ALLSTAR tasks of IRAF \citep{Tody1993}, in a similar way to the procedure described in \citet{Thomas2008}. The optical data were calibrated against the Pan-STARRS catalogue\footnote{See \href{http://archive.stsci.edu/panstarrs/search.php}{http://archive.stsci.edu/panstarrs/search.php}.} \citep{Chambers2016}, while for the near-infrared (NIR) bands, photometric calibration was performed against the 2MASS catalogue \citep{Skrutskie2006AJ}, resulting in a typical absolute accuracy of 0.04~mag in $g'r'i'z'$, 0.06~mag in  $JH$ and 0.08~mag in $K$. 

The 76 cm Katzman Automatic Imaging Telescope located at the Lick Observatory \citep[KAIT;][]{Filippenko2001} obtained $20\times60$~second exposures in the \textit{clear} band \citep[similar to $R$; see][]{Li2003}, starting $\sim$9.04 hours after the BAT trigger \citep{GCN30582}. All images were reduced and co-added using a custom pipeline \citep{Ganeshalingam2010, Stahl2019}, whereafter PSF photometry was performed on the co-added image using DAOPHOT. Several nearby stars were chosen from the Pan-STARRS1 catalogue for flux calibration, with their magnitudes transformed into Landolt magnitudes following Eq. 6 of \cite{Tonry2012}. The optical afterglow of GRB 210731A was clearly detected in the co-added image. 

Images in the SDSS $g$, $r$, and $z$ filters were obtained at a single epoch 1.18~days post-trigger with the acquisition camera of the X-shooter spectrograph, mounted on the ESO Very Large Telescope (VLT) UT3 (Melipal). Reduction was carried out using standard procedures. For the $z$-band image, a fringe correction was applied, using a template fringe pattern provided by the observatory. We also observed the afterglow at two epochs in the $r$ and $z$ bands with the Nordic Optical Telescope (NOT) equipped with the ALFOSC imager. The images were reduced following standard procedures including subtraction of a master bias and correction with sky flats. Magnitudes were measured using aperture photometry, and photometric calibration was carried out against the Pan-STARRS catalogue. 

\begin{figure*}
\centering
\input{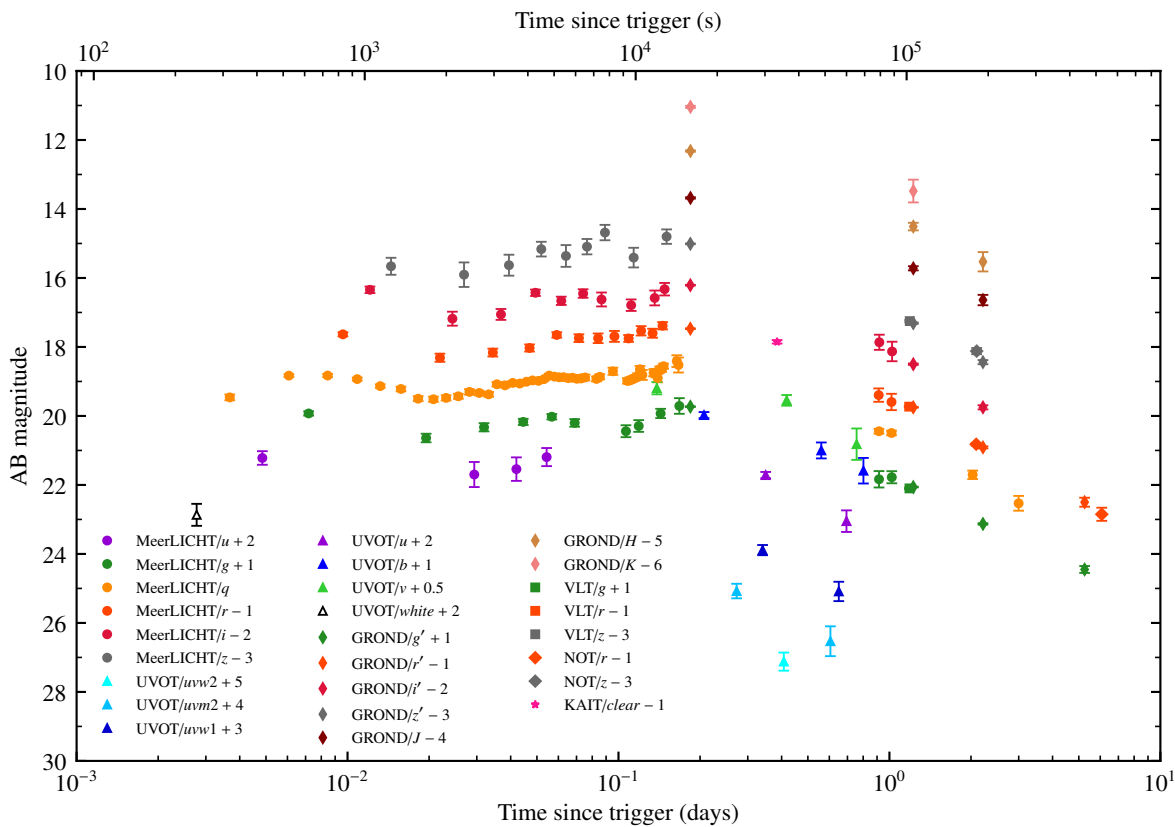}
  \caption{Combined UV/optical/NIR photometry of GRB 210731A. We only show detections where all magnitudes are in the AB system and have not been corrected for Galactic extinction. Times are relative to the \Swift{}/BAT trigger time.}
     \label{fig:shifted_photometry}
\end{figure*}

We show all UV/optical/NIR photometry separated by instrument and filter in Fig. \ref{fig:shifted_photometry}.

\subsection{Radio observations}
We obtained three epochs of radio continuum observations with the MeerKAT radio telescope \citep{Jonas2016} in the L band (1.4~GHz) through director's discretionary time (DDT) proposal DDT-20120810-SD-01 (PI de Wet). Each observation had a total integration time on source of 0.78~hours, using J1939--6342 as the flux and bandpass calibrator and J1924--2914 as the gain calibrator. All data were reduced using the oxkat pipeline\footnote{See \href{https://github.com/IanHeywood/oxkat/blob/master/README.md}{https://github.com/IanHeywood/oxkat/blob/master/README.md} and references therein. } \citep{oxkat}. No radio afterglow was detected at 1.4~GHz across the three epochs at 10.8, 34.1 and 59.7~days post-trigger. The RMS noise was $\approx14$~\si\micro Jy at the $1\sigma$-level in each image. We take upper limits on the afterglow flux as three times the RMS noise. 

Radio continuum observations were also obtained with the \textit{Karl G. Jansky }Very Large Array \citep[JVLA; ][]{Perley2011} in the C and X bands (centred on 6 and 10~GHz) through DDT proposals 21B-333 and 21B-342 (PI de Wet) at four epochs spanning 18.2 to 118~days post-trigger. The total integration time per observation was 0.44~hours in each band, with 3C286 used as the flux and bandpass calibrator and J1924--2914 as the complex gain calibrator. We performed preliminary imaging on the pipeline-calibrated measurements sets using standard Common Astronomy Software Applications \citep[CASA;][]{McMullin2007} procedures and detected the radio afterglow to GRB 210731A in all four epochs at 10~GHz and in all but the first epoch at 6~GHz. 

The first epoch National Radio Astronomy Observatory (NRAO) -calibrated measurement set failed to pass internal quality thresholds for science usability so we chose to calibrate manually from the raw data to obtain more accurate flux measurements. We used CASA version 5.8.0 and performed imaging with the task \verb|tclean| and flux measurements with the task \verb|imfit|. We obtained satisfactory results with our X-band calibration, but the C-band calibration contained persistent phase errors as a result of de-correlation problems during observing, and the measured flux of all sources in the field was substantially lower than in subsequent epochs. Despite these problems, imaging on the re-calibrated data showed a radio source at the afterglow position, in contrast with preliminary imaging. We therefore adopted the following approach to calculate the flux of the afterglow. We identified six point-like sources present at all four epochs in the C band and measured their fluxes. For each source a flux correction factor could then be calculated between the first epoch flux and the flux in each subsequent epoch. The correcting factors ranged from 1.2 to 2.2, with a mean value of 1.77 and standard deviation of 0.26. No obvious trends in the correcting factor were found as a function of source brightness or offset from the image centre across all epochs, so we took the mean correcting factor as the flux correcting factor to apply to the measured afterglow flux in the first epoch image. We incorporate the standard deviation of the correction factor as an additional source of systematic uncertainty in the flux measurement from the first epoch in the C band. 

The third epoch X-band measurement set also had major phase issues, so we performed the same procedure as for the C-band first epoch data, calibrating the raw data and determining a mean flux correcting factor to apply to our afterglow measurement. Only two sources were used in determining the mean correcting factor as the X-band images had far fewer sources than the C-band images. The correcting factors ranged from 1.3 to 1.6, with a mean value of 1.47 and standard deviation of 0.1.

All X-ray, optical, and radio flux measurements associated with GRB 210731A are presented in Table \ref{tab:photometry}.

\section{Afterglow temporal and spectral analysis}\label{sec:results}
We interpret our combined multi-wavelength data in the framework of synchrotron radiation emitted by electrons accelerated to a power-law distribution in energy behind the forward shock, with $N(\gamma)\propto\gamma^{-p}$ for $\gamma>\gamma_\mathrm{min}$, with $\gamma_\mathrm{min}$ being the minimum Lorentz factor of electrons in the distribution and $p$ being the electron energy index, which we assume to be bounded between 2 and 3, though values of less than 2 have been suggested in the literature \citep{Dai2001,Panaitescu2002}. The synchrotron spectra are characterised by power-law segments that join at a number of break frequencies, namely the synchrotron self-absorption frequency $\nu_{sa}$, the characteristic synchrotron frequency $\nu_m$ corresponding to emission from $\gamma_{\rm min}$ electrons, and the cooling frequency $\nu_c$. The orderings of the spectral breaks depend on the hydrodynamic evolution of the forward shock, which is described by the \citet{Blandford1976} spherical self-similar solution of an adiabatic relativistic blast wave expanding into a cold medium with a circumburst density profile varying as a power-law with radius: $n(r)=n_0r^{-k}$. We consider two density profiles: the constant $k=0$ case corresponding to an interstellar-medium-like density profile; and the $k=2$ case corresponding to a stellar wind from a massive star progenitor. The synchrotron forward shock model is described in \citet{Sari1998}, \citet{Chevalier2000}, and \citet{Granot2002}, and we follow the convention $F_{\nu}(t) \propto t^{\alpha}\nu^{\beta}$ throughout. 

\subsection{Optical/X-ray temporal evolution} \label{sec:temporal_evolution}
\begin{figure*}
\centering
\input{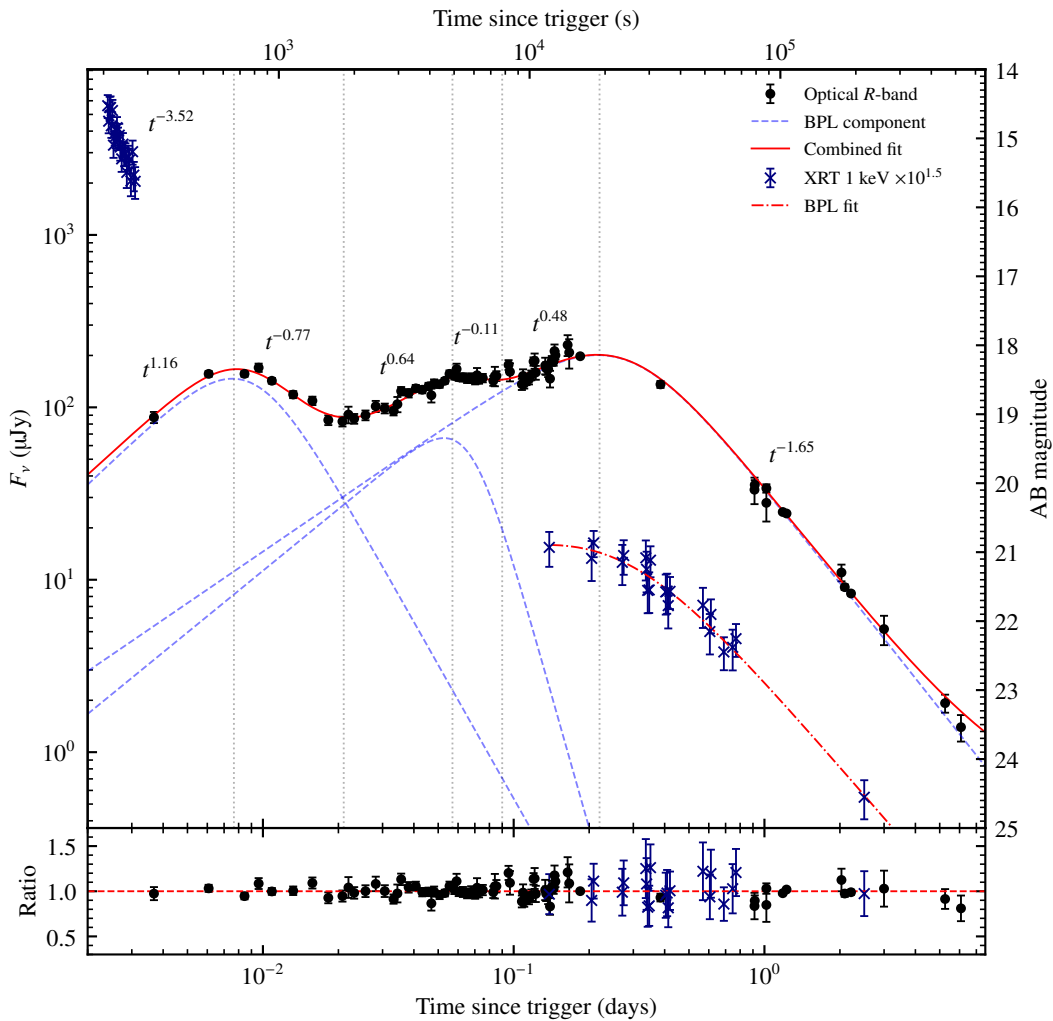}
  \caption{Composite $R$-band light curve and X-ray 1~keV light curve for GRB 210731A. For each rising and decaying segment of the optical light curve, we show the power-law slope ($t^\alpha$) as an indicator of the steepness of the light curve between each pair of adjacent vertical dotted lines. We also show the fit comprising the sum of three BPL components and a constant term equal to the $r'$-band host galaxy flux measured by GROND at 285 days (solid red line), along with each individual BPL component (dashed blue lines). For the X-ray light curve, we indicate the steepness of the WT-mode segment and we show the BPL fit to the PC-mode data as a dashed-dotted red line. The results of the X-ray and optical fits are presented in Table \ref{tab:temporal_fit}. The ratio of observed flux to fitted flux is shown in the lower panel.}
     \label{fig:temporal_fit}
\end{figure*}

The most striking feature of our GRB 210731A dataset is the three peaks in our early-time optical data (see Fig. \ref{fig:shifted_photometry}). To characterise this light curve further, we created a composite $R$-band light curve by combining our $q$-, $r$-, and $R$-band data since they are the most well-sampled optical bands and also have similar central wavelengths. We also included the KAIT \textit{clear} flux measurement since it is calibrated to the $R$ band. We used an optical spectral index of $\beta_\mathrm{opt}=-0.81\pm0.05$ derived from the first GROND epoch (see Sect. \ref{sec:spectral_evolution}) to transform the data to an $R$-band central wavelength of 700 nm\footnote{For direct comparison with the sample in \cite{Li2012} and \cite{Liang2013}.}. The composite $R$-band light curve (see Fig. \ref{fig:temporal_fit}) exhibits three distinct peaks occurring within the first 0.3~days of the GRB trigger, each with rising and decaying segments of varying steepness and smoothness. We investigate the nature of the optical peaks in Sect. \ref{sec:discussion}. After the last peak at ${\sim}0.2$~days, the light curve entered a final declining phase until the last optical observation at 6.2~days post-trigger. 

We follow two approaches to fit the data: first we fitted a single power-law to each rising and decaying segment\footnote{We determine the boundary between each segment by eye. These are shown as vertical dotted lines in Fig. \ref{fig:temporal_fit}.} directly in order to get an indication of the steepness of each segment (we use these in Sect. \ref{sec:e_inj}); then we performed an empirical fit as in \citet{Li2012} by decomposing the light curve into separate components, each of which may arise from different emission sites or physical mechanisms. Since we have three clear peaks in our light curve, we employed a model that comprises the sum of three broken power-law (BPL) components \citep{Beuermann1999,Price2001,Zeh2004}, each characterised by a normalising flux level, $F_0$, rise and decay indices, $\alpha_1$ and $\alpha_2$, break time, $t_\mathrm{b}$ , and break smoothness, $\omega$, according to
\begin{equation} \label{eq:BPL}
F(t) = F_0\left[\left(\frac{t}{t_\mathrm{b}}\right)^{-\alpha_1\omega}+\left(\frac{t}{t_\mathrm{b}}\right)^{-\alpha_2\omega}\right]^{-1/\omega}.
\end{equation}
If $\alpha_1$ is positive and $\alpha_2$ is negative, the light curve peaks at a time, $t_\mathrm{p}$, between rising and decaying segments; $t_\mathrm{b}=t_\mathrm{p}$ in Eq. \ref{eq:BPL}. We also include a constant term to account for the host-galaxy $r'$-band brightness of $24.7\pm0.2$~mag measured at 285~days by GROND. Considering values of 1, 3, 5 and 9 for $\omega$, we find that a smoother break with a value of 1 produces a fit with a $\chi_\mathrm{r}^2$ value closer to 1. Our fit allows us to compare the temporal evolution in each optical band, which we do in Sect. \ref{sec:spectral_evolution}.

Examining the X-ray light curve, the WT-mode data within the first 300 seconds shows a steep decline with $\alpha_\mathrm{X}=-3.52\pm0.36$ and a photon index that hardened from 3.2 to 2.2, as taken from the online Burst Analyser. The most likely explanation is high latitude prompt emission, as the temporal and spectral indices agree broadly with the $\alpha=-2+\beta$ curvature effect relation \citep{Kumar2000,Willingale2010}. It is unfortunate that we have no X-ray data during the time of the first two optical peaks, rendering a direct comparison between the two bands unfeasible. There is, however, X-ray data from ${\sim}0.13$~days onwards starting when the optical light curve was rising to its final peak. The X-ray light curve started in a shallow decay or plateau phase before steepening, which coincided with the final decaying phase in the optical. We fit a BPL according to Eq. \ref{eq:BPL} to determine the break time and temporal slopes, fixing the break smoothness at 1, 3, 5, or 9. Each fit had a similar reduced $\chi^2_r$ value (${\sim}0.4$); therefore, we employed a break smoothness of $\omega=1$ to match the value used in the optical fit, though this results in a pre-break index that is poorly constrained. The $R$-band and 1 keV light curves are presented in Fig. \ref{fig:temporal_fit} along with their fits, and we present the results of the fits in Table \ref{tab:temporal_fit}.

\begin{table*}
\caption{Parameters derived from fits to the composite $R$-band light curve and X-ray 1~keV light curve, as shown in Fig. \ref{fig:temporal_fit}. The $R$-band light curve was fitted with the sum of three BPLs, while the X-ray light curve was fitted with a single BPL. }\label{tab:temporal_fit}
\centering
\begin{tabular}{ccccc}
\hline
\hline
 & $\alpha_1$ & $\alpha_2$ & $t_\mathrm{p}$ (days) & $\chi^2$/dof\\
\hline
Optical BPL 1 & $1.39\pm0.36$ & $-2.58\pm0.75$ & $0.0088\pm0.0013$ & 1.32\\
Optical BPL 2 & $1.19\pm0.62$ & $-5.16\pm2.58$ & $0.066\pm0.007$ & -\\
Optical BPL 3 & $0.99\pm0.16$ & $-1.84\pm0.04$ & $0.27\pm0.01$ & -\\
X-ray 1 keV BPL & $0.44\pm0.62$ & $-1.69\pm0.19$ & $0.26\pm0.09$ & 0.39\\

\hline
\end{tabular}
\end{table*}

\subsection{Achromatic optical/X-ray spectral evolution} \label{sec:spectral_evolution}
\begin{figure}
\centering
\input{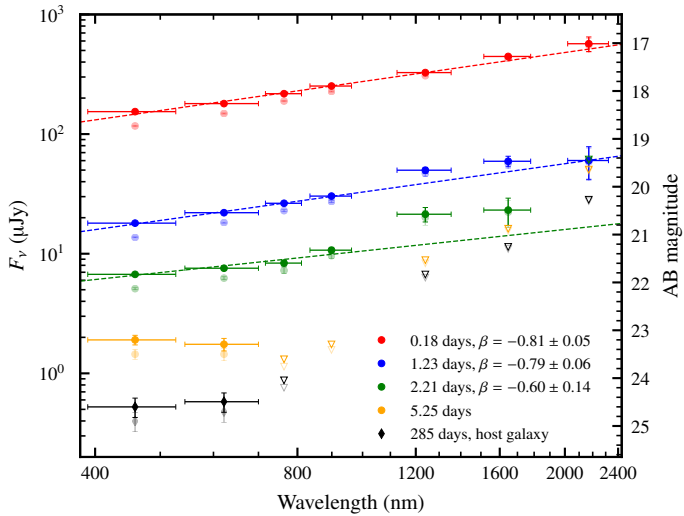}
  \caption{Power-law fits to the GROND optical/NIR data at three epochs, corrected for a Galactic extinction of $A_V=0.24$~mag. The uncorrected magnitudes are shown in a lighter shade below each corrected data point. We also show the fourth epoch, which only had detections in the $g'$ and $r'$ bands, as well as the detections (shown as diamonds) and limits from deep observations of the host galaxy at~285 days. Upper limits are shown as upside-down triangles. }
     \label{fig:GROND_SED}
\end{figure}

We now investigate if there is evidence for spectral evolution in the optical data. In Fig. \ref{fig:GROND_SED} we show the optical spectral energy distributions (SEDs) formed using data from the first three of five GROND epochs corrected for a Galactic extinction of $A_\mathrm{V}=0.24$ mag in the direction of the GRB \citep{Schlafly2011}. We fitted the data with power-laws in frequency, with the first epoch yielding a spectral index of $\beta_\mathrm{opt}=-0.81\pm0.05$. There does not appear to be substantial spectral evolution between the first two epochs at 0.184 and 1.225 days, particularly in the optical $g'$, $r'$, $i'$, and $z'$ bands. It is unclear why there is excess emission in the near infrared $J$, $H$, and $K$ bands during the second and third epochs. A possible explanation is that there is contaminating emission from the host galaxy, in which case we would expect the light curves to flatten towards a constant value. The observed decline to below detection levels at 5.25 and 285 days appears to rule out this possibility. It could also be the case that there is an additional unaccounted-for source of systematic photometric error.  

\begin{figure}
\centering
\input{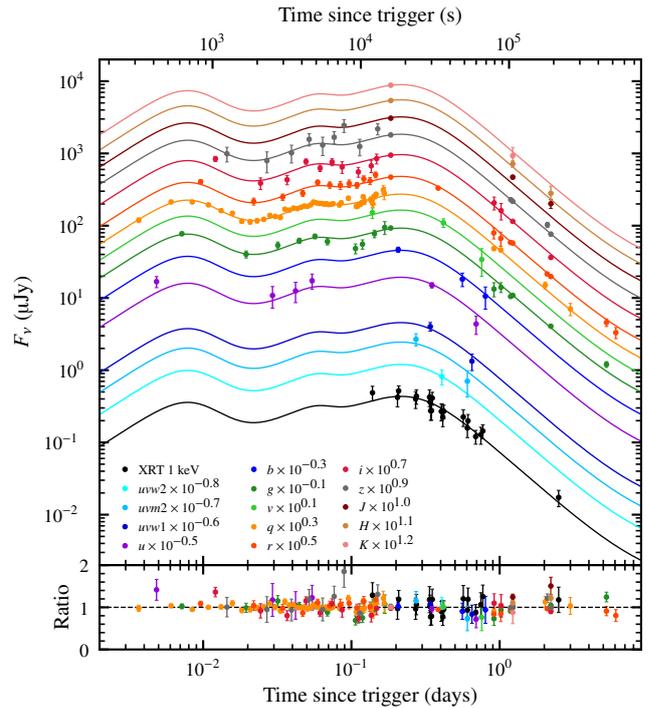}
  \caption{Fits to the data in each optical band using the best-fit model to the composite $R$-band light curve in Fig. \ref{fig:temporal_fit}. We also fitted this model to the X-ray PC-mode data. The lower panel presents the ratio of measured flux to model flux.}
     \label{fig:fit_filter}
\end{figure}

For our early-time data, we took our composite $R$-band light curve fit (Fig. \ref{fig:temporal_fit}) and fitted the flux in each of the optical bands with this model, which amounts to shifting the $R$-band fit light curve vertically until it fits the data in a given band. The spectral slope in the blue and UV bands ($u$ through $uvw2$) was steeper than in the optical owing to Galactic extinction and damping by Ly$\alpha$ absorption at the redshift of the burst. We therefore shifted the MeerLICHT $u$-band data to the UVOT $u$ band using an approximate spectral index of $\beta_\mathrm{UV}\approx-4$ measured from the UVOT/$u$, $uvw1$, $uvm1$, and $uvw2$ data. Figure \ref{fig:fit_filter} shows that the data in each of the UV and optical bands is reasonably well fitted by the $R$-band light curve. Deviations from the fit at earlier times ($<0.2$~days) are visible in the $u$, $g$, $i$, and $z$ bands but they do not appear statistically significant - only 7.5\% of the UV/optical/NIR data points deviate by $2\sigma$ or more from each model fit. Overall the optical evolution appears achromatic and consistent with a constant optical spectrum, which points towards a hydrodynamical rather than spectral origin to the complex early-time optical light curve.  

As mentioned in Sect. \ref{sec:XRT}, insufficient X-ray photons were collected during the PC-mode observations to create time-sliced spectra. The photon index from the online Burst Analyser was fairly constant during PC mode, however, with a mean value of $\Gamma_\mathrm{X}=1.84\pm0.27$. This is indicative of insubstantial spectral evolution in the X-ray band during PC mode. The X-ray spectral index of $\beta_\mathrm{X}=-1.00\pm0.11$ from our spectral fit in Sect. \ref{sec:XRT} (with a mean photon arrival time of 0.42~days) is steeper than the first GROND epoch (at 0.18~days) spectral index of $\beta_\mathrm{opt}=-0.81\pm0.05$, by $\Delta\beta=0.19\pm0.12$. This difference is suggestive of a spectral break lying between the X-ray and optical bands. If the break was the cooling break, we would expect $\Delta\beta=0.5$, however. This discrepancy could be explained by the fact that the cooling break is always smooth \citep{Uhm2014} or that $\nu_c$ may lie near to either the optical or X-ray bands. We investigate this further in Sect. \ref{sec:closure_relations}.

\subsection{Closure relation analysis}\label{sec:closure_relations}
Within the synchrotron forward shock model the `closure relations' relate the spectral index $\beta$ with the temporal index $\alpha$ in the convention $F_{\nu}\propto t^{\alpha}\nu^{\beta}$ and depend on the physical regime, spectral regime, and external medium density profile \citep{Zhang2004,Zhang2006}. The closure relations can be adapted to describe a variety of alternative scenarios to the standard self-similar deceleration phase, including the post-jet break scenario, whether there is energy injection involved, the reverse shock crossing phase, or the Newtonian/non-relativistic phase \citep[see the comprehensive review in][]{Gao2013}. It is important to note that the spectral breaks are inherently smooth, so that a transitioning spectral break or spectral break near to an observing band may define a `grey zone' where the $\alpha-\beta$ relations are not strictly satisfied \citep{Zhang2006,Uhm2014}.

If we assume that the final declining phase of our optical and X-ray light curves arises from standard forward shock emission in the slow cooling regime ($\nu_m<\nu_c$), which is usually the case at later times, we can determine which spectral regime and density profile best fits our data by performing a closure relation analysis. Our X-ray and optical data both have negative spectral slopes, with fluxes that decline with increasing frequency. There are two spectral regimes that give rise to negative spectral slopes: $\nu_c<\nu$ (Regime I) with $\beta=-p/2$, or $\nu_m<\nu<\nu_c$ (Regime II) with $\beta=(1-p)/2$. For the optical spectral index of $\beta_\mathrm{opt}=-0.81\pm0.05$ derived from the first GROND epoch we would have $p=1.62\pm0.10$ or $p=2.62\pm0.10$ in Regime I and II, respectively. The decay index for both a wind or interstellar medium (ISM) environment in Regime I is $\alpha=(2-3p)/4$ resulting in $\alpha=-0.72\pm0.08$ with $p=1.62\pm0.10$, too shallow for the observed decay rate of $\alpha_\mathrm{opt}=-1.84\pm0.04$ from our three BPL-component fit. In Regime II, we have $\alpha=3(1-p)/4$ or $\alpha=(1-3p)/4$ for an ISM or wind density profile, respectively. With $p=2.62\pm0.10$ we get $\alpha=-1.22\pm0.08$ or $\alpha=-1.72\pm0.08$ for the ISM or wind profiles. Clearly, the observed optical temporal index of $\alpha_\mathrm{opt}=-1.84\pm0.04$ is most consistent with the optical spectral index in Regime II for a wind profile. 

In a wind medium the cooling break moves to higher frequencies as $t^{1/2}$, with light curves that are shallower by $\Delta\alpha=0.25$ above $\nu_c$. With the cooling break between the optical and X-rays one could therefore expect the X-ray light curve to decline more slowly than the optical light curve. Our fits to the X-ray and optical light curves in Sect. $\ref{sec:temporal_evolution}$ result in $\Delta\alpha=0.15\pm0.19$, which agrees within uncertainties with the predicted difference of $\Delta\alpha=0.25$. We also see, however, that this $\Delta\alpha$ value is consistent with a zero difference within $1\sigma$ and is supported by the fact that the composite $R$-band light curve provides a good fit to the X-ray light curve, as shown in Fig. \ref{fig:fit_filter}. It is therefore not possible to conclusively say whether $\nu_c$ lies between the optical and X-ray bands from our data, though we note that the slightly different temporal and spectral (see Sect. \ref{sec:spectral_evolution}) indices between the two bands does hint at this possibility. In summary, our optical and X-ray data can be accommodated within the standard closure relations in a wind medium. 

\subsection{Broadband SED evolution}\label{sec:opt_radio}
The optical and X-ray data alone can only place weak constraints on the location of $\nu_m$ and the peak flux of the evolving synchrotron spectrum. Our late-time radio observations, which probe the low-frequency end of the synchrotron spectrum, can provide valuable constraints on the location of $\nu_m$ and the peak flux, and can therefore lead to an estimation of the intrinsic blast wave parameters (see Sect. \ref{sec:modelling}). 

Our three epochs of L-band (at 1.4~GHz) observations all yielded non-detections, whereas our four epochs of C-band and X-band data yielded detections spanning 18 to 118~days post-trigger. The flux across the first two epochs at 18.2 and 34.2~days was fairly constant in both the C and X bands, which is consistent with the predicted evolution of $t^0$ for the spectral ordering $\nu_{sa}<\nu<\nu_m$ in a wind medium undergoing slow cooling. The spectral slope between the first epoch C- and X-band detections of $\beta_{6-10~\mathrm{GHz}}=0.57\pm0.27$ is also close to the predicted value of $\nu^{1/3}$ for the same spectral segment. The subsequent decline in flux across the last two epochs can be interpreted as the passage of $\nu_m$ through 6 and 10~GHz, whereafter both bands lie in Regime II of the synchrotron spectrum in which the flux declines with time and the spectrum declines with increasing frequency. This is seen in our last two epochs where the C-band flux is in fact brighter than the X-band flux. At 34.2 days we have quasi-simultaneous flux measurements at 1.4, 6 and 10~GHz (see Fig.~\ref{fig:radio_SED}). The spectral index between the C and X bands is $\beta_{6-10~\mathrm{GHz}}=0.69\pm0.23$, which is closer to the optically thin spectral slope of $\nu^{1/3}$ than the synchrotron self-absorbed slope of $\nu^2$. Based on the C-band detection and L-band upper limit, we place a lower limit on the spectral slope of $\beta_{1.4-6~\mathrm{GHz}}>0.84$. It therefore may be the case that synchrotron self-absorption is responsible for the non-detections at 1.4~GHz, since our L-band limit is consistent with a $\nu^2$ spectrum. In that case, the self-absorption frequency could lie between 1.4 and 10~GHz at 34.2~days. 

In a wind medium, $\nu_m$ moves to lower frequencies as $t^{-3/2}$ with the corresponding peak flux of the synchrotron spectrum declining as $t^{-1/2}$ for the spectral break ordering $\nu_{sa}<\nu_m<\nu_c$. From our GROND SED at ${t=0.184}$~days (Fig. \ref{fig:GROND_SED}), we know that $\nu_m$ lies below the $K$ band with a peak flux greater than the measured $K$-band flux of 555~\si\micro Jy. If we assume that $\nu_m$ passes through the radio X band (10~GHz) at 34.2~days with a peak flux of ${\sim}250$~\si\micro Jy, we would have expected $\nu_m$ to be at $1.48\times10^{12}$ Hz with a peak flux of 1320~\si\micro Jy at 1.22~days, the time of our second GROND epoch. This frequency lies below the $K$-band frequency of $1.38\times10^{14}$~Hz, as expected. The spectral index between this expected peak flux value and the measured $K$-band flux value at 1.22~days results in $\beta\approx-0.7$, which is in agreement with the measured optical spectral index of $\beta_\mathrm{opt}=-0.79\pm0.06$ at this time. We also note that the X-ray to optical $R$-band spectral index at ${\sim}0.38$~days ($\beta_\mathrm{opt,X}\approx-0.95$) is between the X-ray-only ( $\beta_\mathrm{X}=-1.00\pm0.11$) and optical-only index ($\beta_\mathrm{opt}=-0.81\pm0.05$), demonstrating that both observing bands can be accommodated via a forward shock model.

With these basic considerations, we attempted to find a first-guess set of blast wave parameters that can explain our data. We have the following assumed constraints: (i) $\nu_m$ passes through 10~GHz at 34.2~days; (ii) the corresponding peak flux at this frequency and time is ${\sim}250$~\si\micro Jy; (iii) $\nu_c$ lies between the optical and X-ray bands at early times (i.e. $\nu_c \approx 10^{17}$~Hz at 0.3~days); and (iv) $\nu_{sa}$ lies between 1.4 and 10~GHz at 34.2~days. 

With these four constraints we can attempt to solve the system of four equations describing the locations of the spectral breaks and their corresponding flux densities in a wind medium, given in Table 2 of \citet{Granot2002}. Our solution given the above constraints results in an unphysical value of $\epsilon_e>1$, which is driven primarily by the requirement that the self-absorption frequency lies between 1.4 and 10~GHz at 34.2~days. Lowering $\nu_{sa}$ to a frequency of ${\sim}10^7$~Hz results in a physical solution for all of the blast wave parameters. Our L-band limits therefore pose a challenge to the interpretation of our multi-wavelength afterglow data. We return to this point in Sect. \ref{sec:L_band}.

\subsection{Early jet-break scenario}\label{sec:jetbreak}
Our optical light curve during the final declining phase had a temporal index of $\alpha_\mathrm{opt}=-1.84\pm0.04$ from the combined fit or $\alpha_\mathrm{opt}=-1.65\pm0.04$ from the direct fit to the late-time data only, which is steep for normal pre-jet break evolution \citep[see Fig. 4 in][]{Wang2018}. An alternative scenario to explain the steep final declining phase in the optical and X-ray light curves is post-jet break decay. If the jet break is due to a purely geometric edge effect \citep{Panaitescu1998}, the light curves within all spectral regimes should steepen by $t^{-3/4}$ for the ISM case and $t^{-1/2}$ for the wind case once the ejecta has slowed down such that the relativistic beaming angle $1/\Gamma$ is greater than the jet half-opening angle $\theta_j$, assuming a top-hat jet. Sideways expansion of a conical jet would result in a steeper jet break decay of approximately $t^{-p}$ in Regimes I and II for an ISM \citep{Rhoads1999,Sari1999b}. 

Considering the edge effect only, a jet break will not change the temporal evolution of the synchrotron spectral break frequencies. If an early jet break occurred we would expect our radio data to show declining light curves that decay as $t^{-1/2}$ in a wind medium under the assumption that $\nu_{sa}<\nu_{6,10~\mathrm{GHz}}<\nu_m$. Taking into account sideways expansion, the evolution of the break frequencies is altered, though we would still expect declining light curves at radio frequencies \citep{Sari1999b}. The rising radio light curves in the C and X bands until ${\sim}34$~days are therefore inconsistent with an early jet break. This implies that the steep optical and X-ray decline is normal pre-jet break decay in a wind medium, supporting our analysis in Sect. \ref{sec:closure_relations}. 

\section{Theoretical modelling}\label{sec:modelling}
We have shown in the previous sections that our X-ray, optical, and radio data after the last optical peak can be reconciled within the synchrotron forward shock model in a wind medium with $p\approx2.6$ if we exclude our L-band limits. We now proceed to find a set of blast wave parameters that can describe our data by employing the smoothly connected power-law spectra outlined in \citet{Granot2002} and fitting for the forward shock parameters $p$, $\epsilon_e$, $\epsilon_B$, $A_\star$, and $E_\mathrm{K,iso}$, where $E_\mathrm{K,iso}$ is the total isotropic-equivalent kinetic energy in the blast wave; $\epsilon_e$ and $\epsilon_B$ are the fractions of shock internal energy given to the electrons and the magnetic fields, respectively; and $A_\star=A/(5\times10^{11}\mathrm{~g~cm}^{-1})$ is the wind density parameter as defined in \citet{Chevalier2000}.  We correct the data for Galactic extinction with $A_V=0.24$~mag. 

We perform a Markov chain Monte Carlo (MCMC) analysis with \texttt{emcee} \citep{emcee} using 512 walkers and 2000 steps, discarding the initial 250 steps as burn-in. The details of our implementation are described in \citet{Laskar2013,Laskar2014}. The host galaxy extinction, $A_\mathrm{V,host}$, is a free parameter in our model. We include the effects of Klein-Nishina (KN) corrections (G.~ McCarthy \& T.~ Laskar in prep) using prescriptions from \citet{Nakar2009} and \citet{Jacovich2021}. We used uniform, uninformative priors flat in log space, and restricted $\epsilon_e+\epsilon_B<1$, although this limit is not reached. We did not include data before the inferred time of the last optical peak at ${\sim}0.3$ days in the modelling, and discuss these data further in Sect.~\ref{sec:discussion}. We also did not include the MeerKAT $1.4$~GHz observations, as we do not expect these to be fit with this model (Sect.~\ref{sec:opt_radio}). We present theoretical  modelling including the L-band limits in Appendix \ref{sec:appendix_modelling}. For completeness, we also include the possibility of a jet break following \citet{Rhoads1999}. We set a lower limit on the jet break time of $t_{\rm jet}\gtrsim34$~days since there is no evidence for an earlier jet break in the data, as discussed in Sect. \ref{sec:jetbreak}.

\begin{figure}
\centering
\input{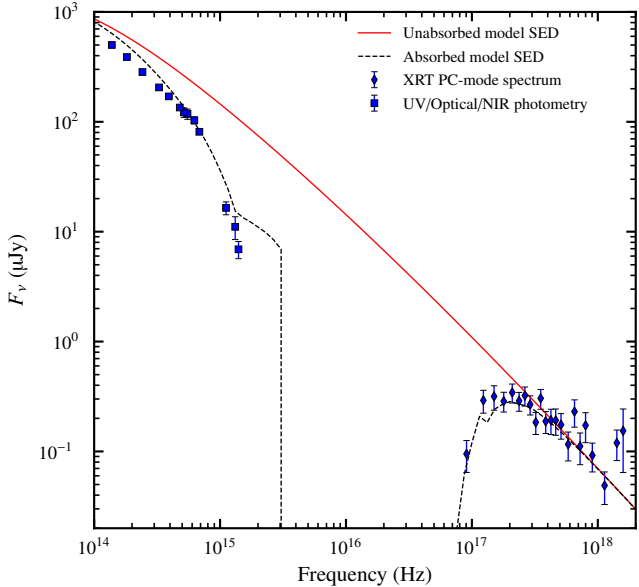}
  \caption{Optical to X-ray SED at 0.3~days along with the highest-likelihood theoretical model SED, both absorbed (dashed line) and unabsorbed (solid line). The optical photometric data points were derived from the light curve fits to each observing band (see Fig. \ref{fig:fit_filter}) through interpolating each fit to 0.3~days. The X-ray PC-mode spectrum had a mean photon arrival time of 0.43~days, so we used the X-ray BPL light curve fit in Fig. \ref{fig:temporal_fit} to determine a correcting factor to shift the spectrum to the expected flux level at 0.3~days.}
     \label{fig:SED}
\end{figure}

\begin{figure*}
\centering
\input{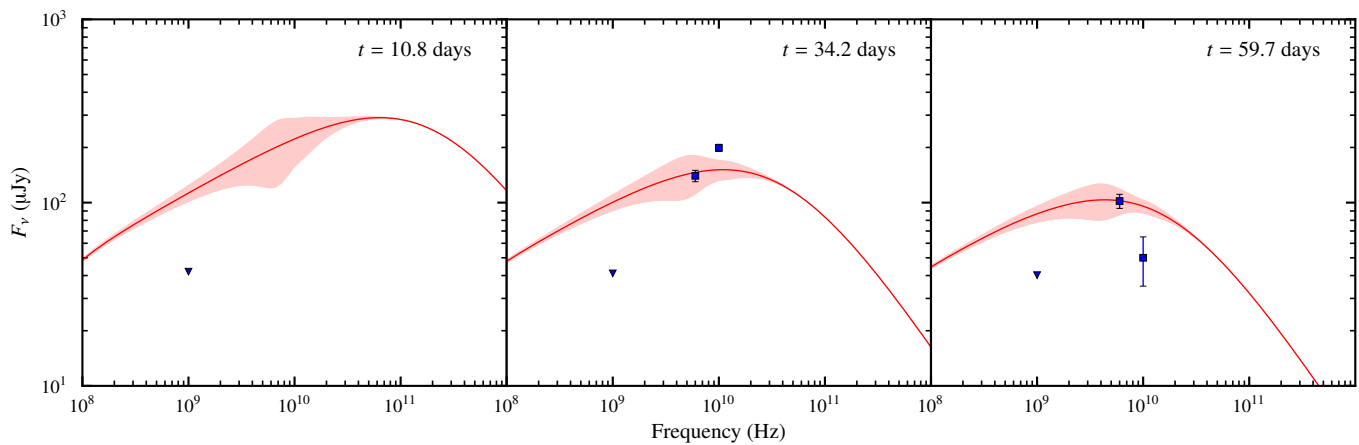}
  \caption{Radio SEDs at the times of the three MeerKAT epochs. We show the highest-likelihood unabsorbed model SED along with the effects of Galactic scintillation at the $1\sigma$ level as derived from the NE2001 model \citep{Cordes2002} for the GRB line of sight through the Milky Way. For the second epoch we show the C- and X-band detections obtained quasi-simultaneously at 34.2~days, while for the third epoch we show the C- and X-band detections obtained 7.4~days after the third MeerKAT epoch. }
     \label{fig:radio_SED}
\end{figure*}

\begin{figure}
\centering
\input{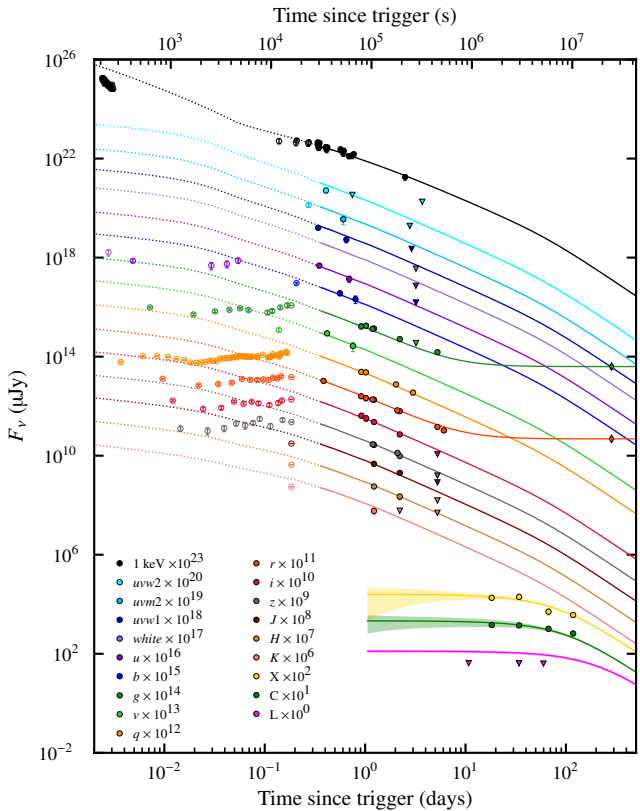}
  \caption{Light curves from our highest-likelihood theoretical model shown for each observing band, spanning X-ray, optical, and radio frequencies. Only data points after 0.3~days were used in the modelling, and we show these as filled-in data points, in contrast to the earlier time data points shown as empty circles. Upper limits are shown as upside-down triangles. The shaded regions surrounding the three radio bands (X, C, and L) represent the effects of Galactic scintillation at the $1\sigma$ level. The optical $g$- and $r$-band model fits plateau towards the measured host-galaxy flux levels at 285~days, shown as diamonds. }
     \label{fig:LC_model}
\end{figure}

\begin{figure*}
\centering
\input{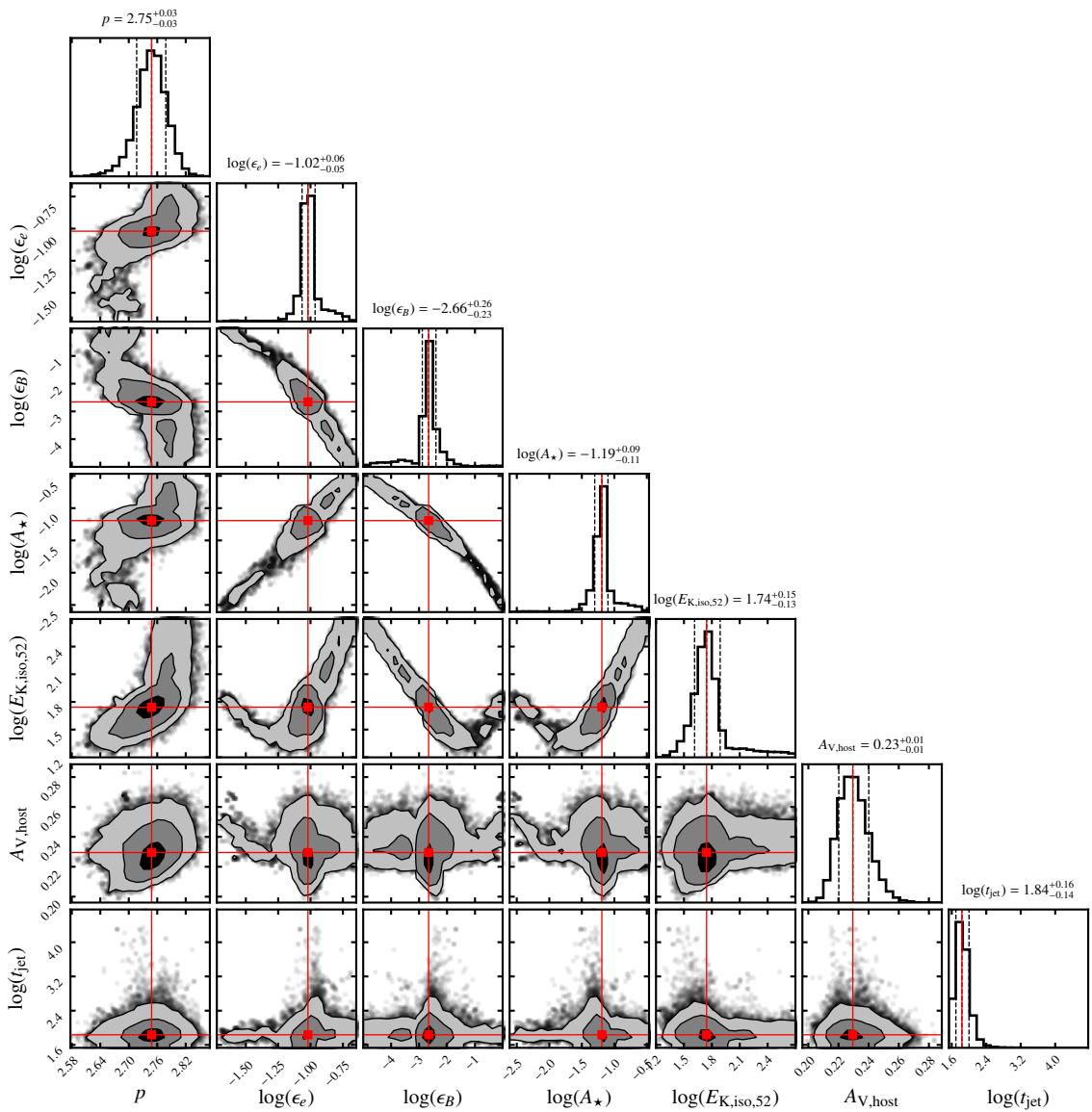}
  \caption{Corner plot showing the marginalised posterior distributions for each model parameter along with the 2D marginalised posterior distributions for each pair of model parameters, from our MCMC analysis. Contours are at the $1\sigma$, $2\sigma$, and $3\sigma$ levels, and the red lines denote the median values derived from the marginalised posterior distributions for each model parameter.}
     \label{fig:corner}
\end{figure*}

The physical parameters for the highest-likelihood model and those derived from the MCMC analysis are presented in Table \ref{tab:MCMC_params}, while the corresponding light curves are presented in Fig. \ref{fig:LC_model}. For these parameters, both inverse Compton and KN effects are important at early times ($\lesssim1$~day), with Compton $Y\approx4$ at ${\sim}0.3$~days. At this time, the relevant spectral break frequencies are located at $\nu_m\approx4\times10^{13}$~Hz, $\nu_c\approx \hat{\nu}_c\approx 10^{16}$~Hz, and $\hat{\nu}_m\approx10^{21}$~Hz, resulting in the spectral ordering $\nu_{\rm opt}<\nu_c\approx\hat{\nu}_c\lesssim\nu_{\rm X}$, where $\hat{\nu}_c$ and $\hat{\nu}_m$ are KN spectral breaks as outlined in \citet{Nakar2009}. The cooling frequency passes through the X-ray band between $\approx0.6$ to 12~days, consistent with the discussion in Sects.~\ref{sec:spectral_evolution} and \ref{sec:closure_relations}. In this regime, the spectral index in the X-rays is expected to be intermediate between $(1-p)/2\approx-0.88$ and $-p/2\approx-1.34$, which is consistent with the observed X-ray spectral index of $\beta_{\rm X}=-1.00\pm0.11$. Taking into account the $1\sigma$ confidence intervals from the MCMC analysis, the derived value of $p=2.75\pm0.03$ is consistent with the value of $p=2.62\pm0.10$ inferred from our closure relation analysis in Sect. \ref{sec:closure_relations}. This model also requires an intrinsic extinction of $A_\mathrm{V,host}\approx0.2$~mag, consistent with the observed UV suppression (Fig.~\ref{fig:SED}). The corner plot in Fig. \ref{fig:corner} shows that there are strong correlations between some pairs of parameters, especially those involving $\epsilon_e$, $\epsilon_B$, $A_\star$, and $t_\mathrm{jet}$. The model over-predicts the $1.4$~GHz flux by a factor of $\approx3$ with respect to our MeerKAT upper limits (Fig.~\ref{fig:radio_SED}); even when taking scintillation  into account, the upper limits are all more than $4\sigma$ below the model flux. We return to this point in Sect.~\ref{sec:L_band}.

Our shock microphysics parameters are fairly typical. Our value of $p=2.75\pm0.03$ is within the $1\sigma$ range of the sample in \citet{Wang2015}, for which they find $p=2.33\pm0.48$. \citet{Santana2014} collect $\epsilon_e$ and $\epsilon_B$ values in the literature and find that $\epsilon_e$ is narrowly distributed across one order of magnitude between ${\sim}0.02$ to 0.6 with a median of value 0.22. Our value of $\approx0.1$ is a normal value within their sample. For the magnetic field equipartition factor, they find a much wider distribution varying across almost 5 orders of magnitude from ${\sim}3.5\times10^{-5}$ to 0.33. Our value of $\epsilon_B\approx0.2\times10^{-2}$ is close to their median value of $1.0\times10^{-2}$. Additionally, \citet{Beniamini2017} use radio light curve peaks to determine the distribution of $\epsilon_e$ and find a value of $\mathrm{log}_{10}\epsilon_e=-0.88\pm0.26$ for a wind medium. Again, our derived valued is consistent with their sample value. 

The model requires a jet break at $t_{\rm jet}\approx64$~days, which is driven by the declining radio light curves after this time. For the highest-likelihood parameters, this corresponds to a jet opening angle of $\theta_{\rm jet}\approx5^{\circ}$. In the absence of a jet break, the model over-predicts the final X-band detection by $\approx3.5\sigma$. However, we note that the evidence in support of a jet break is fairly weak and this inferred opening angle should be interpreted with caution. For a limit of $t_{\rm jet}\gtrsim118$~days (the last radio detection), the highest-likelihood model yields a lower limit on the opening angle of $\theta_{\rm jet}\gtrsim6^{\circ}$. The corresponding beaming correction of ${f_b=(1-\cos\theta_{\rm jet})\gtrsim5.48\times10^{-3}}$ implies constraints on the true $\gamma$-ray and kinetic energy of $E_{\gamma}\gtrsim7.07\times10^{49}$~erg and $E_\mathrm{K}\gtrsim3.45\times10^{51}$~erg, respectively, where we have used the maximum-likelihood (ML) estimate from Table \ref{tab:MCMC_params} for $E_\mathrm{K,iso}$. 

\begin{table}
\caption{Parameters derived from our multi-wavelength theoretical modelling of the afterglow data after 0.3~days. We show both the highest-likelihood model parameters from a maximum-likelihood (ML) estimation and the median values along with their corresponding $1\sigma$ confidence intervals from the MCMC marginalised posterior distributions presented in Fig. \ref{fig:corner}. The beaming-corrected prompt $\gamma$-ray and kinetic energies are given in the lower panel of the table, where we have placed a lower limit on the opening angle of the jet based on a limit of $t_\mathrm{jet}\gtrsim118$ days. }\label{tab:MCMC_params}
\centering
\begin{tabular}{ccc}
\hline
\hline
 Parameter & ML estimate & MCMC results \\
\hline
$p$ & 2.75 & $2.75\pm0.03$ \\
$\epsilon_e$ & $9.7\times10^{-2}$ & $9.6^{+1.3}_{-1.0}\times10^{-2}$ \\
$\epsilon_B$ & $1.7\times10^{-3}$ & $2.2^{+1.9}_{-0.9}\times10^{-3}$ \\
$A_\star$ & $7.1\times10^{-2}$ & $6.4^{+1.6}_{-1.5}\times10^{-2}$ \\
$E_\mathrm{K,iso}$ ($10^{52}$~erg) & 63 & $69^{+32}_{-19}$ \\
$A_\mathrm{V,host}$ (mag) & 0.22 & $0.23\pm0.01$ \\
$t_\mathrm{jet}$ (days) & 64 & $55^{+23}_{-14}$ \\
\hline
$\theta_\mathrm{jet}$ (deg) & $\gtrsim6$ & - \\
$E_\gamma$ (erg) & $\gtrsim7.07\times10^{49}$ & - \\
$E_\mathrm{K}$ (erg) & $\gtrsim3.45\times10^{51}$ & - \\
\hline
\end{tabular}
\end{table}

\section{Discussion}\label{sec:discussion}
The optical light curve of GRB 210731A is unusual for showing three distinct peaks of similar brightness within the first five hours of the GRB trigger. Multiple peaks in optical light curves have been observed before \citep[e.g. GRBs 060904B, 080928, 100621A, 100814A, 100901A;][]{Klotz2008,Rossi2011,Greiner2013,Nardini2014,Laskar2015}. We investigate a number of explanations proposed in the literature for peaks and re-brightenings in afterglow light curves, including the passage of a spectral break, flaring behaviour, secondary jets, an off-axis viewing angle, energy injection into the forward shock, and the onset of afterglow. We also consider the implications of our L-band upper limits. 

\subsection{The nature of the optical re-brightenings}\label{sec:peaks}
\subsubsection{Passage of a spectral break}
The passage of the spectral break associated with the peak of the synchrotron spectrum ($\nu_m$) through the optical bands could in principle give rise to a peak in the optical light curves. Since $\nu_m$ moves to lower frequencies with increasing time, we would expect the spectral index to transition from a positive to negative value over the rise and fall of the light curve. We have shown in Sect. \ref{sec:spectral_evolution} and in Fig. \ref{fig:fit_filter}, however, that our optical evolution is achromatic, ruling out a spectral break origin to any of the peaks. 

\subsubsection{Flaring behaviour}
The peaks in the early light curve are too smooth and long-lived to be due to flares. Hence, the observed structure in the optical light curve cannot be due to any of the mechanisms that are typically invoked to explain flares, such as late-time central engine activity, density fluctuations, or reverse shock emission.

\subsubsection{Off-axis viewing angle}
It is possible to obtain a rising light curve if the GRB jet is viewed from an angle outside the cone of the jet (i.e. $\theta_\mathrm{obs}\gtrsim\theta_\mathrm{jet}$; \citealt{Granot2002a}). The peak of the light curve corresponds to the time when the Lorentz factor of the jet is $\sim1/\theta_\mathrm{obs}$, whereafter the light curve evolves in a post-jet break manner. Our radio data and theoretical modelling does not support an early jet break however, as discussed in Sects. \ref{sec:jetbreak} and \ref{sec:modelling}. Although it would be possible to explain a single peak in the light curve through viewing angle effects, it is difficult to interpret all three light curve peaks within such a scenario. Due to the relativistic beaming effect, one would also expect to observe negligible prompt $\gamma$-ray emission when the viewing angle is outside the jet cone, resulting in an orphan afterglow \citep{Rhoads1999,Granot2002a,Zou2007}. The prompt $\gamma$-ray observations of GRB 210731A therefore do not support an off-axis viewing angle interpretation. 

\subsubsection{Two-component jet model}
The two-component jet model \citep{Peng2005} has been invoked to explain chromatic behaviour and late-time re-brightenings observed in a number of GRB afterglows \citep{Berger2003,Racusin2008,Filgas2011,Nicuesa2011,Kann2018}. In this model a fast, narrow inner jet powers the prompt emission and early afterglow emission, while a slower, wider jet powers the late-time afterglow evolution, with both jets viewed on-axis. This model was preferred by \citet{Liang2013} to explain the re-brightenings observed in their sample of optical afterglows, in which the deceleration of the slow jet explains the re-brightening peaks, analogous to the onset peaks. They claim that the similar rising and decaying indices of the onset and re-brightening peaks supports this interpretation. Furthermore, they find that the properties of the re-brightenings are not correlated with the prompt emission properties (contrary to the onset peaks), and so they are likely independent emission components. The fact that the GRB 210731A optical light curve shows three distinct peaks appears to rule out the two-component jet model. A three-component jet might be able to explain the three light curve peaks, but we consider it beyond the scope of this work. 

\subsubsection{Energy injection}\label{sec:e_inj}
The most straightforward explanation for the optical light curve evolution is energy injection into the forward shock. Energy injection, or a refreshed shock, has been invoked to explain the plateaus and shallow decay segments seen in X-ray \citep{Campana2005,Vaughan2006,Nousek2006,Zhang2006,Liang2007,Evans2009} and optical light curves \citep{Mangano2007,Greiner2009,Swenson2013}. In this framework, the blast wave energy increases with time during the energy injection period, rather than remaining constant. Two physical mechanisms have been proposed: the first is a long-lasting central engine that continuously injects a Poynting flux into the blast wave as in the case of a spin-down millisecond magnetar \citep{Dai1998,Zhang2001a}, where the central engine luminosity is described as a power law in time with $L(t)=L_0(t/t_0)^{-q}$. The injected energy $E_\mathrm{inj}$ is essentially constant when $q\geq1$, while the total energy in the blast wave can only increase significantly with time when $q<1$. Once the injected energy begins to exceed the original energy in the blast wave, the total energy will scale as $E_\mathrm{tot}\propto t^{1-q}$. The second case could arise from an impulsive central engine injection episode, producing a stratified ejecta distribution where the ejecta mass above a certain Lorentz factor $\Gamma$ is described as a power law, for example $M(>\Gamma)\propto\Gamma^{-s}$. The energy in a shell with Lorentz factor $\Gamma$ is added to the blast wave when the blast wave bulk Lorentz factor has slowed down to $\Gamma$, so that the energy in the blast wave scales as $E(>\Gamma)\propto \Gamma^{1-s}$ \citep{Rees1998,Sari2000}. For both cases the blast wave scaling laws can be derived and applied to the synchrotron spectra to obtain closure relations that depend on either $q$ or $s$. Both forms of energy injection can be cast in an equivalent form, and simple relations between $q$ and $s$ can be derived \citep{Zhang2006}. From a closure relation analysis with energy injection alone, it is therefore impossible to distinguish energy injection from a long-lasting central engine or from injection of a stratified ejecta.

The $R$-band light curve segments between the first and final peaks have temporal indices of $\alpha=[-0.77,0.64,-0.11,0.48]$ (from the direct fit to each segment in Fig. \ref{fig:temporal_fit}). Adopting the $q$-formalism, we can determine the energy injection index $q$ for each segment by making use of the closure relations for a wind medium in the slow cooling regime \citep{Zhang2006}. We make the assumption that the $R$ band remained in the spectral regime satisfying $\nu_m<\nu_R<\nu_c$ during all four segments of energy injection, which is valid since our data supports achromatic optical evolution. In this regime, $\alpha=\frac{(2-2p)-(p+1)q}{4}$, so employing $p=2.75$ from our theoretical modelling we derive values of $q=[-0.11,-1.62,-0.82,-1.45]$ for each light curve segment after the first peak. The energy increase during a time period from $t_0$ to $t_1$ is calculated as 
\begin{equation}\label{eq:e_increase}
    E_\mathrm{K,iso,1} = E_\mathrm{K,iso,0}\left(\frac{t_1}{t_0}\right)^{1-q}. 
\end{equation}
Assuming that the blast wave kinetic energy evolves according to Eq. \ref{eq:e_increase} during each of the four power law segments, we can determine the energy at the time of the first peak, $E_\mathrm{K,iso,0}$, using 
\begin{equation}
    E_\mathrm{K,iso,f} = E_\mathrm{K,iso,0}\left(\frac{t_1}{t_0}\right)^{1-q_1}\left(\frac{t_2}{t_1}\right)^{1-q_2}\left(\frac{t_3}{t_2}\right)^{1-q_3}\left(\frac{t_4}{t_3}\right)^{1-q_4}
\end{equation}
along with the start and end times of each segment\footnote{We use the vertical dashed lines in Fig. \ref{fig:temporal_fit} as the start and end times of each segment. These have values of $t_0=0.0077$, $t_1=0.021$, $t_2=0.057$, $t_3=0.09$, and $t_4=0.22$~days.}, our values calculated for $q$ above, and a final blast wave energy of $E_\mathrm{K,iso,f}=6.3\times10^{53}$~erg from our theoretical modelling. We find that the blast wave energy at the time of the first optical peak is equal to $7.3\times10^{50}$~erg, smaller by a factor of ${\sim}1000$ compared to the final kinetic energy, indicative of substantial energy injection.  

\citet{Laskar2015} argue that the significant X-ray and optical re-brightenings seen in a sample of GRB afterglows are best explained by the stratified ejecta model, since energy injection from a spinning-down millisecond magnetar should not lead to a significant increase in the blast wave energy (i.e. $q\geq 1$). They also exclude fall-back accretion onto a black hole as the theoretically-predicted accretion rate is insufficient to power plateaus or re-brightenings. In the stratified ejecta formalism, there is a gap between the initial blast wave shell and the fast outer shell of the stratified ejecta that is moving with some maximum Lorentz factor, $\Gamma_\mathrm{max}$. As the initial shell slows down, the stratified ejecta deposits energy into the blast wave until the slowest shell moving with Lorentz factor $\Gamma_\mathrm{min}$ has deposited its energy, whereafter the afterglow evolves following the standard framework. From their study of a sample of afterglows exhibiting later-time re-brightenings, \citet{Laskar2015} showed that a large amount of the kinetic energy deposited into the blast wave comes from the slowest-moving ejecta. They also find that the GRBs with significant energy injection have low radiative efficiencies, consistent with the prompt $\gamma$-ray emission being produced by the fastest-moving ejecta and a large amount of kinetic energy being carried by the slower ejecta. With our value of $E_{\gamma,\mathrm{iso}}=1.29\times10^{52}$~erg and $E_\mathrm{K,iso}=63\times10^{52}$~erg, we calculate a radiative efficiency using $\eta\equiv E_{\gamma,\mathrm{iso}}/(E_{\gamma,\mathrm{iso}}+E_\mathrm{K,iso})$ of $\eta\approx 0.02$, a low radiative efficiency consistent with significant energy injection. We also note that the energy released in $\gamma$ rays is $\approx18$ times greater than the kinetic energy in the blast wave derived at the time of the first optical peak. This is unphysical if the kinetic energy at that time is the true energy reservoir from which the prompt emission is drawn, and adds support to the final kinetic energy being the true energy reservoir.  

\subsubsection{Afterglow onset and initial Lorentz factor, $\Gamma_0$} \label{sec:onset}
The synchrotron forward shock model predicts a smooth onset peak in optical light curves as the expanding blast wave is slowed down by the circumburst medium \citep{Sari1999,Kobayashi2007}.  Such onset peaks are not uncommon in early optical afterglows and have been studied by a number of authors  \citep{Zhang2003,Molinari2007,Xue2009,Liang2010,Liang2013}. For an ISM density profile, the deceleration time (or onset peak time) is most sensitive to the bulk Lorentz factor of the ejecta and depends weakly on other parameters. Onset peaks can therefore provide a valuable way of constraining the initial Lorentz factor $\Gamma_0$ of the GRB blast wave. For a non-ISM density profile, however, the dependence of the deceleration time on other parameters becomes stronger. 

The first optical detection associated with GRB 210731A was made in the UVOT $white$ filter starting 210~seconds post-trigger, with subsequent detections showing a steady rise to a smooth peak around 700~seconds. Thereafter, the light curve entered a declining phase before starting to rise steadily again at ${\sim}$1700~seconds. \citet{Li2012} and \citet{Liang2013} define an onset peak as a smooth hump peaking within one hour post-trigger that is followed by a normal power-law decay component. By comparing our onset peak with the sample of 38 onset peaks in \citet{Liang2013}, we can assess how likely it is that our first peak is indeed the onset of afterglow. From our combined \textit{R}-band fit (Fig. \ref{fig:temporal_fit} and Table \ref{tab:temporal_fit}) we derive a number of additional properties from each BPL component. These include the peak flux ($F_\mathrm{p}$), peak time ($t_\mathrm{p}$), full width at half maximum ($w$), peak $R$-band luminosity ($L_\mathrm{R,p}$), and the isotropic energy release in the interval $[t_\mathrm{p}/5,5t_\mathrm{p}]$, as shown in Table \ref{tab:temporal_fit_params}. 
\begin{table*}
\caption{Parameters derived from each BPL component of our $R$-band light curve fit.}\label{tab:temporal_fit_params}
\centering
\begin{tabular}{ccccc}
\hline
\hline
 $F_\mathrm{p}$ ($10^{-12}$~erg~cm$^{-2}$~s$^{-1}$) & $t_\mathrm{p}$ (s) & $w$ (s) & $L_\mathrm{R,p}$ ($10^{45}$~erg~s$^{-1}$) & $E_\mathrm{R,iso}$ ($10^{48}$~erg)\\
\hline
$1.92\pm0.27$ & $760\pm111$ & $956$ & $15.69\pm2.22$ & $8.28$ \\
$0.75\pm0.27$ & $5704\pm645$ & $5385$ & $6.11\pm2.21$ & $16.96$ \\
$2.62\pm0.10$ & $23161\pm1179$ & $39824$ & $21.41\pm0.78$ & $464.53$ \\
\hline
\end{tabular}
\end{table*}

\citet{Liang2013} find that the onset peak times of their sample span a range of 30-3000 seconds, with typical rising and decaying indices of 1.5 and $-1.15$ in ranges of [0.3,4] and [$-1.8$,$-0.6$], respectively. Our peak time of $760\pm111$ seconds and rising index of $1.39\pm0.36$ are typical values within this sample, but our decaying index of $-2.58\pm0.75$ from BPL 1 in Table \ref{tab:temporal_fit} is steeper than average, though it does fall within the sample range within uncertainty.  It should be noted that the decaying slope of our first peak is not well determined owing to the uncertain contribution of the second BPL component at this time. Furthermore, the authors derive a number of empirical relations between pairs of properties of onset peaks (see their Figs. 7 and 9): the width of a peak is strongly correlated with the peak time; the \textit{R}-band peak luminosity is anti-correlated with the rest-frame time; and the peak luminosity and energy are correlated with the isotropic $\gamma$-ray energy. We find that our measured values in Table \ref{tab:temporal_fit_params} agree closely with their empirical relations, lending support to the onset peak claim. 

Under the assumption that GRB 210731A occurred in a stellar wind medium ($k=2$), we calculate the initial Lorentz factor following \citet{Zhang2018} as
\begin{equation}\label{eq:gamma0}
\Gamma_0\simeq 1.3^{1/4}\left( \frac{3E_\mathrm{K,iso}(1+z)}{8\pi A c^3 t_\mathrm{dec}}\right)^{1/4} \simeq 120 t_\mathrm{dec}^{-1/4}\left(\frac{1+z}{2}\right)^{1/4}E_{52}^{1/4}A_\star^{-1/4},
\end{equation}
where after the last equality, $t_\mathrm{dec}$, is measured in seconds, $E_{52}$ in units of $10^{52}$~erg, and $A_\star$ is the wind density parameter as described previously. Equation \ref{eq:gamma0} depends on the assumption of an impulsive fireball, that is, the thin shell regime. The isotropic-equivalent kinetic energy of the blast wave $E_\mathrm{K,iso}$ can be inferred from theoretical modelling of the afterglow, which we do in Sect. \ref{sec:modelling}. However, the value calculated in Sect. \ref{sec:modelling} was only applicable to the late-time light curve, so we employ the value for $E_\mathrm{K,iso}$ of $7.3\times10^{50}$~erg at the time of the first optical peak calculated in Sect. \ref{sec:e_inj}.  Using the $A_\star$ value from our theoretical modelling and the peak time of our first peak, we calculate an initial Lorentz factor of $\Gamma_0\approx24$.  As mentioned previously, the dependence of the deceleration time on other parameters is stronger for a non-ISM density profile. This is also the case if there is energy injection during deceleration, which may be the case during our early optical observations (Sect. \ref{sec:e_inj}). We therefore caution that our calculation of $\Gamma_0$ is an estimate. 

\subsection{Suppressed L-band flux}\label{sec:L_band}
Our highest-likelihood theoretical model can provide an adequate fit to all of our late-time data except for the L band, where the model over-predicts the flux by a factor of $\approx3$ compared to our MeerKAT upper limits. We did not expect our model to fit the L-band data since we found that requiring the synchrotron self-absorption frequency to lie above the L band at 34.2~days resulted in an unphysical value of $\epsilon_e>1$. We therefore need to consider an additional source of opacity at these observing frequencies to explain our L-band limits.

One possible source of additional opacity is a thermal electron population within the GRB shock front. \citet{Ressler2017} modelled afterglow spectra and light curves while considering the effect of such a population and find that it has two effects on the spectra: an excess of flux near the peak synchrotron frequency of the thermal electrons that fades with time as the electrons cool; and additional opacity in the optically thick portion of the spectrum compared to the case with only non-thermal electrons. The latter effect is consistent with a higher self-absorption frequency by a factor of 10-100. It could therefore be the case that our suppressed L-band flux points towards a population of thermal electrons. We leave the detailed modelling including a thermal population of electrons to a future work.  

\section{Conclusion}
GRB 210731A was a long-duration burst discovered by \Swift{}/BAT. Observations with the optical telescope MeerLICHT starting 286~seconds post-trigger found an unusual optical light curve evolution with three peaks of similar brightness within the first 4.3~hours; afterwards, the burst entered a declining phase. We find that the early optical evolution is consistent with a constant optical spectrum, pointing towards a hydrodynamical origin. A closure relation analysis based on the optical SED and temporal decay after the last peak showed a preference for a stellar wind environment, consistent with the long GRB duration and therefore a massive star origin. We find that the first optical peak can be explained as the onset of afterglow, while energy injection into the forward shock from a stratified ejecta is a natural explanation for the two subsequent re-brightenings. We estimate that the blast wave kinetic energy increased by a factor of ${\sim}1000$ from the first optical peak to the last peak. Detailed theoretical modelling of the optical, X-ray, and radio data after the last optical peak at ${\sim}0.3$~days resulted in typical blast wave and shock microphysics parameters. Our MeerKAT L-band upper limits could not be reconciled with our model, however, possibly implying a thermal electron population within the shocked region that provided an additional source of opacity. Future multi-wavelength modelling of GRB afterglows, especially at millimetre and radio frequencies, will shed light on the electron distribution in GRB shocks.  

\begin{acknowledgements}
The MeerLICHT consortium is a partnership between Radboud University, the University of Cape Town, the Netherlands Organisation for Scientific Research (NWO), the South African Astronomical Observatory (SAAO), the University of Oxford, the University of Manchester and the University of Amsterdam, in association with and, partly supported by, the South African Radio Astronomy Observatory (SARAO), the European Research Council and the Netherlands Research School for Astronomy (NOVA). We acknowledge the use of the Inter-University Institute for Data Intensive Astronomy (IDIA) data intensive research cloud for data processing. IDIA is a South African university partnership involving the University of Cape Town, the University of Pretoria and the University of the Western Cape. SdW and PJG are supported by NRF SARChI Grant 111692. Part of the funding for GROND (both hardware and personnel) was generously granted from the Leibniz-Prize to G. Hasinger (DFG grant HA 1850/28-1) and by the Th\"uringer Landessternwarte Tautenburg. AVF is grateful for financial assistance from the Christopher R. Redlich Fund and numerous individual donors. KAIT and its ongoing operation were made possible by donations from Sun Microsystems, Inc., the Hewlett-Packard Company, AutoScope Corporation, Lick Observatory, the U.S. NSF, the University of California, the Sylvia \& Jim Katzman Foundation, and the TABASGO Foundation.  Research at Lick Observatory is partially supported by a generous gift from Google. This work made use of data supplied by the UK \Swift{} Science Data Centre at the University of Leicester. This publication made use of the python package \verb|corner.py| \citep{corner}.  
\end{acknowledgements}

\bibliographystyle{aa}
\bibliography{refs}

\clearpage
\onecolumn

\begin{appendix} 
\section{Theoretical modelling with MeerKAT L-band limits}\label{sec:appendix_modelling}
\begin{figure}
\centering
\input{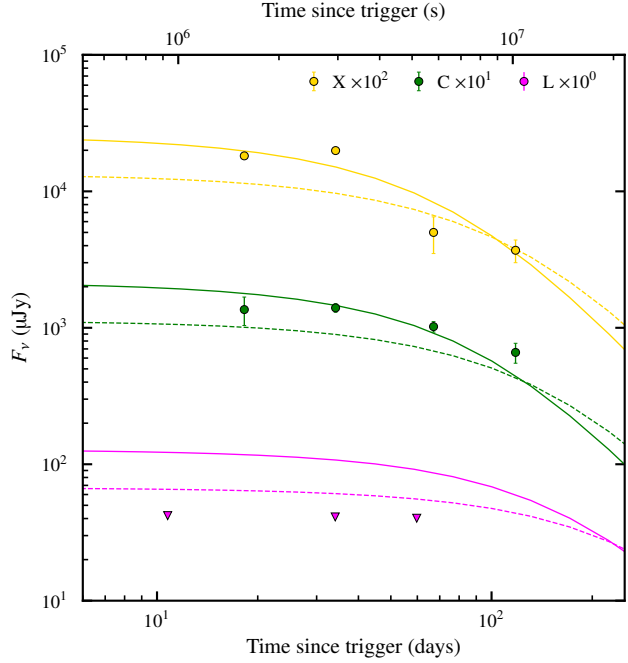}
  \caption{Highest-likelihood model for the fit that includes the L-band data (dashed lines) and the fit that excludes the L-band data (solid lines). The fit to the optical and X-ray light curves is similar in both models. }
     \label{fig:LC_model_MKT}
\end{figure}

\begin{figure*}
\centering
\input{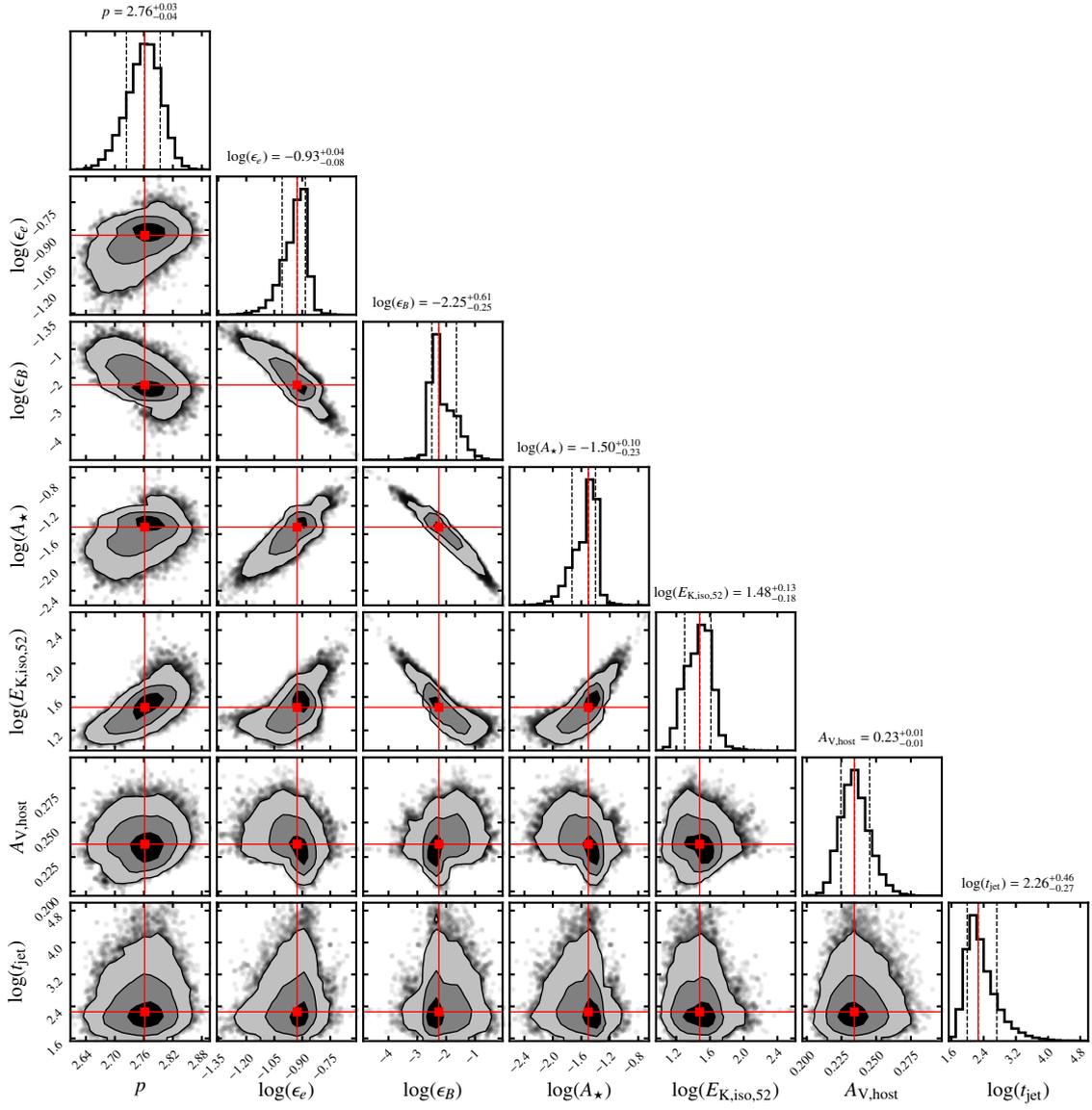}
  \caption{Same as Fig. \ref{fig:corner}, but for the fit including the L-band upper limits.}
     \label{fig:corner_MKT}
\end{figure*}

We repeated the theoretical modelling following the procedure outlined in Sect. \ref{sec:modelling}, but including the MeerKAT L-band limits. The radio light curves from the highest-likelihood model are shown in Fig. \ref{fig:LC_model_MKT}, and the MCMC parameter distributions are presented in Fig. \ref{fig:corner_MKT} and Table \ref{tab:MCMC_MKT_params}. 

Including the MeerKAT L-band limits leads to a poorer fit. This is demonstrated by the highest-likelihood light curve under-predicting the C- and X-band fluxes while still over-predicting the L-band flux, and the actual likelihood value of this fit being lower than for the fit excluding the MeerKAT limits. We measured a maximum log likelihood of 236.9 for the fit without the limits vs 225.1 for the fit including the MeerKAT limits. As we expected, including the MeerKAT data does not lead to a model with a steep enough spectral index to accommodate the non-detections. 

\begin{table}
\caption{Same as Table \ref{tab:MCMC_params}, but for the fit including the L-band upper limits.}
\label{tab:MCMC_MKT_params}
\centering
\begin{tabular}{ccc}
\hline
\hline
 Parameter & ML estimate & MCMC results \\
\hline
$p$ & 2.76 & $2.76^{+0.03}_{-0.04}$ \\
$\epsilon_e$ & $1.2\times10^{-2}$ & $1.2^{+0.1}_{-0.2}\times10^{-1}$ \\
$\epsilon_B$ & $3.1\times10^{-3}$ & $5.7^{+17.7}_{-2.5}\times10^{-3}$ \\
$A_\star$ & $4.1\times10^{-2}$ & $3.1^{+0.8}_{-1.3}\times10^{-2}$ \\
$E_\mathrm{K,iso}$ ($10^{52}$ erg) & 39 & $30^{+11}_{-10}$ \\
$A_\mathrm{V,host}$ (mag) & 0.22 & $0.23\pm0.01$ \\
$t_\mathrm{jet}$ (days) & 125 & $186^{+37}_{-87}$ \\

\hline
\end{tabular}
\end{table}

\section{Flux measurements}\label{sec:appendix_photometry}
Table \ref{tab:photometry} contains all the X-ray, optical, and radio observations used in this work.

\begin{longtable}{ccccccc}
\caption{X-ray/optical/radio flux measurements of GRB 210731A.}\label{tab:photometry}\\
\hline
\hline
\makecell{$\Delta t$ \\ (days)}  & Telescope & Band/Filter & \makecell{Frequency \\(Hz)} & \makecell{Flux \\(\si\micro Jy)} & \makecell{Uncertainty \\(\si\micro Jy)} & \makecell{Detection? \\ ($1=$ yes)} \\
\hline
\endfirsthead
\caption{Continued.} \\
\hline
\hline
\makecell{$\Delta t$ \\ (days)}  & Telescope & Band/Filter & \makecell{Frequency \\(Hz)} & \makecell{Flux \\(\si\micro Jy)} & \makecell{Uncertainty \\(\si\micro Jy)} & \makecell{Detection? \\ ($1=$ yes)} \\
\hline
\endhead
\hline
\endfoot
\hline
\endlastfoot
0.00241 & \Swift{}/XRT & 1 keV & 2.42e+17 & 177.589 & 27.077 & 1 \\
0.00243 & \Swift{}/XRT & 1 keV & 2.42e+17 & 144.408 & 21.527 & 1 \\
0.00245 & \Swift{}/XRT & 1 keV & 2.42e+17 & 173.501 & 26.453 & 1 \\
0.00247 & \Swift{}/XRT & 1 keV & 2.42e+17 & 173.423 & 27.678 & 1 \\
0.00249 & \Swift{}/XRT & 1 keV & 2.42e+17 & 135.870 & 20.254 & 1 \\
0.00251 & \Swift{}/XRT & 1 keV & 2.42e+17 & 166.523 & 25.390 & 1 \\
0.00254 & \Swift{}/XRT & 1 keV & 2.42e+17 & 104.861 & 16.356 & 1 \\
0.00257 & \Swift{}/XRT & 1 keV & 2.42e+17 & 122.493 & 18.260 & 1 \\
0.00259 & \Swift{}/XRT & 1 keV & 2.42e+17 & 118.424 & 17.654 & 1 \\
0.00262 & \Swift{}/XRT & 1 keV & 2.42e+17 & 132.423 & 20.190 & 1 \\
0.00264 & \Swift{}/XRT & 1 keV & 2.42e+17 & 115.389 & 17.201 & 1 \\
0.00267 & \Swift{}/XRT & 1 keV & 2.42e+17 & 119.856 & 19.577 & 1 \\
0.00270 & \Swift{}/XRT & 1 keV & 2.42e+17 & 102.343 & 17.113 & 1 \\
0.00273 & \Swift{}/XRT & 1 keV & 2.42e+17 & 87.334 & 13.938 & 1 \\
0.00276 & \Swift{}/XRT & 1 keV & 2.42e+17 & 106.299 & 16.965 & 1 \\
0.00279 & \Swift{}/XRT & 1 keV & 2.42e+17 & 94.213 & 15.036 & 1 \\
0.00283 & \Swift{}/XRT & 1 keV & 2.42e+17 & 93.477 & 13.935 & 1 \\
0.00287 & \Swift{}/XRT & 1 keV & 2.42e+17 & 72.925 & 13.765 & 1 \\
0.00290 & \Swift{}/XRT & 1 keV & 2.42e+17 & 89.665 & 14.310 & 1 \\
0.00294 & \Swift{}/XRT & 1 keV & 2.42e+17 & 85.993 & 14.046 & 1 \\
0.00298 & \Swift{}/XRT & 1 keV & 2.42e+17 & 64.979 & 11.965 & 1 \\
0.00302 & \Swift{}/XRT & 1 keV & 2.42e+17 & 96.119 & 15.340 & 1 \\
0.00305 & \Swift{}/XRT & 1 keV & 2.42e+17 & 70.237 & 13.596 & 1 \\
0.00309 & \Swift{}/XRT & 1 keV & 2.42e+17 & 64.675 & 13.529 & 1 \\
0.13866 & \Swift{}/XRT & 1 keV & 2.42e+17 & 0.507 & 0.116 & 1 \\
0.20459 & \Swift{}/XRT & 1 keV & 2.42e+17 & 0.438 & 0.114 & 1 \\
0.20839 & \Swift{}/XRT & 1 keV & 2.42e+17 & 0.538 & 0.093 & 1 \\
0.27115 & \Swift{}/XRT & 1 keV & 2.42e+17 & 0.415 & 0.109 & 1 \\
0.27459 & \Swift{}/XRT & 1 keV & 2.42e+17 & 0.455 & 0.103 & 1 \\
0.33652 & \Swift{}/XRT & 1 keV & 2.42e+17 & 0.441 & 0.115 & 1 \\
0.33905 & \Swift{}/XRT & 1 keV & 2.42e+17 & 0.378 & 0.100 & 1 \\
0.34243 & \Swift{}/XRT & 1 keV & 2.42e+17 & 0.285 & 0.075 & 1 \\
0.34705 & \Swift{}/XRT & 1 keV & 2.42e+17 & 0.287 & 0.076 & 1 \\
0.35107 & \Swift{}/XRT & 1 keV & 2.42e+17 & 0.427 & 0.087 & 1 \\
0.40418 & \Swift{}/XRT & 1 keV & 2.42e+17 & 0.280 & 0.073 & 1 \\
0.40918 & \Swift{}/XRT & 1 keV & 2.42e+17 & 0.283 & 0.074 & 1 \\
0.41349 & \Swift{}/XRT & 1 keV & 2.42e+17 & 0.233 & 0.061 & 1 \\
0.42004 & \Swift{}/XRT & 1 keV & 2.42e+17 & 0.282 & 0.060 & 1 \\
0.56853 & \Swift{}/XRT & 1 keV & 2.42e+17 & 0.234 & 0.061 & 1 \\
0.60579 & \Swift{}/XRT & 1 keV & 2.42e+17 & 0.165 & 0.043 & 1 \\
0.61160 & \Swift{}/XRT & 1 keV & 2.42e+17 & 0.207 & 0.046 & 1 \\
0.68902 & \Swift{}/XRT & 1 keV & 2.42e+17 & 0.125 & 0.027 & 1 \\
0.74695 & \Swift{}/XRT & 1 keV & 2.42e+17 & 0.133 & 0.035 & 1 \\
0.77010 & \Swift{}/XRT & 1 keV & 2.42e+17 & 0.150 & 0.032 & 1 \\
2.50026 & \Swift{}/XRT & 1 keV & 2.42e+17 & 0.018 & 0.005 & 1 \\
\hline
0.00368 & MeerLICHT & $q$ & 5.169e+14 & 59.71 & 4.40 & 1 \\
0.00485 & MeerLICHT & $u$ & 7.889e+14 & 74.48 & 13.03 & 1 \\
0.00607 & MeerLICHT & $q$ & 5.169e+14 & 106.67 & 3.93 & 1 \\
0.00719 & MeerLICHT & $g$ & 6.246e+14 & 97.28 & 5.38 & 1 \\
0.00844 & MeerLICHT & $q$ & 5.169e+14 & 106.67 & 3.93 & 1 \\
0.00962 & MeerLICHT & $r$ & 4.789e+14 & 128.24 & 7.09 & 1 \\
0.01086 & MeerLICHT & $q$ & 5.169e+14 & 97.28 & 3.58 & 1 \\
0.01210 & MeerLICHT & $i$ & 3.919e+14 & 167.50 & 13.88 & 1 \\
0.01320 & MeerLICHT & $q$ & 5.169e+14 & 80.91 & 3.73 & 1 \\
0.01447 & MeerLICHT & $z$ & 3.276e+14 & 124.75 & 28.72 & 1 \\
0.01575 & MeerLICHT & $q$ & 5.169e+14 & 74.48 & 4.12 & 1 \\
0.01823 & MeerLICHT & $q$ & 5.169e+14 & 57.55 & 3.71 & 1 \\
0.01950 & MeerLICHT & $g$ & 6.246e+14 & 50.59 & 5.59 & 1 \\
0.02076 & MeerLICHT & $q$ & 5.169e+14 & 56.50 & 3.64 & 1 \\
0.02193 & MeerLICHT & $r$ & 4.789e+14 & 68.55 & 7.58 & 1 \\
0.02311 & MeerLICHT & $q$ & 5.169e+14 & 58.62 & 3.78 & 1 \\
0.02438 & MeerLICHT & $i$ & 3.919e+14 & 77.27 & 14.23 & 1 \\
0.02564 & MeerLICHT & $q$ & 5.169e+14 & 61.38 & 3.96 & 1 \\
0.02691 & MeerLICHT & $z$ & 3.276e+14 & 100.01 & 32.24 & 1 \\
0.02818 & MeerLICHT & $q$ & 5.169e+14 & 69.19 & 5.10 & 1 \\
0.02936 & MeerLICHT & $u$ & 7.889e+14 & 47.87 & 15.87 & 1 \\
0.03061 & MeerLICHT & $q$ & 5.169e+14 & 67.30 & 4.34 & 1 \\
0.03189 & MeerLICHT & $g$ & 6.246e+14 & 67.30 & 7.44 & 1 \\
0.03318 & MeerLICHT & $q$ & 5.169e+14 & 64.87 & 3.58 & 1 \\
0.03438 & MeerLICHT & $r$ & 4.789e+14 & 78.71 & 7.97 & 1 \\
0.03559 & MeerLICHT & $q$ & 5.169e+14 & 84.73 & 4.68 & 1 \\
0.03684 & MeerLICHT & $i$ & 3.919e+14 & 86.30 & 12.72 & 1 \\
0.03811 & MeerLICHT & $q$ & 5.169e+14 & 82.42 & 4.55 & 1 \\
0.03942 & MeerLICHT & $z$ & 3.276e+14 & 128.24 & 35.43 & 1 \\
0.04071 & MeerLICHT & $q$ & 5.169e+14 & 87.91 & 4.05 & 1 \\
0.04199 & MeerLICHT & $u$ & 7.889e+14 & 55.47 & 17.37 & 1 \\
0.04321 & MeerLICHT & $q$ & 5.169e+14 & 86.30 & 3.97 & 1 \\
0.04450 & MeerLICHT & $g$ & 6.246e+14 & 77.99 & 6.46 & 1 \\
0.04577 & MeerLICHT & $q$ & 5.169e+14 & 90.37 & 4.16 & 1 \\
0.04696 & MeerLICHT & $r$ & 4.789e+14 & 88.72 & 8.17 & 1 \\
0.04815 & MeerLICHT & $q$ & 5.169e+14 & 93.76 & 5.18 & 1 \\
0.04939 & MeerLICHT & $i$ & 3.919e+14 & 154.18 & 12.78 & 1 \\
0.05064 & MeerLICHT & $q$ & 5.169e+14 & 92.90 & 4.28 & 1 \\
0.05189 & MeerLICHT & $z$ & 3.276e+14 & 197.71 & 38.24 & 1 \\
0.05312 & MeerLICHT & $q$ & 5.169e+14 & 97.28 & 3.58 & 1 \\
0.05430 & MeerLICHT & $u$ & 7.889e+14 & 76.56 & 18.33 & 1 \\
0.05548 & MeerLICHT & $q$ & 5.169e+14 & 106.67 & 4.91 & 1 \\
0.05675 & MeerLICHT & $g$ & 6.246e+14 & 89.54 & 6.60 & 1 \\
0.05801 & MeerLICHT & $q$ & 5.169e+14 & 103.76 & 4.78 & 1 \\
0.05919 & MeerLICHT & $r$ & 4.789e+14 & 125.90 & 9.28 & 1 \\
0.06037 & MeerLICHT & $q$ & 5.169e+14 & 101.87 & 4.69 & 1 \\
0.06149 & MeerLICHT & $i$ & 3.919e+14 & 124.75 & 12.64 & 1 \\
0.06275 & MeerLICHT & $q$ & 5.169e+14 & 101.87 & 5.63 & 1 \\
0.06398 & MeerLICHT & $z$ & 3.276e+14 & 164.45 & 46.95 & 1 \\
0.06522 & MeerLICHT & $q$ & 5.169e+14 & 100.01 & 5.53 & 1 \\
0.06766 & MeerLICHT & $q$ & 5.169e+14 & 100.93 & 6.51 & 1 \\
0.06893 & MeerLICHT & $g$ & 6.246e+14 & 75.86 & 7.69 & 1 \\
0.07020 & MeerLICHT & $q$ & 5.169e+14 & 98.18 & 5.43 & 1 \\
0.07143 & MeerLICHT & $r$ & 4.789e+14 & 115.88 & 11.74 & 1 \\
0.07275 & MeerLICHT & $q$ & 5.169e+14 & 99.09 & 5.48 & 1 \\
0.07403 & MeerLICHT & $i$ & 3.919e+14 & 151.37 & 16.73 & 1 \\
0.07527 & MeerLICHT & $q$ & 5.169e+14 & 100.93 & 5.58 & 1 \\
0.07651 & MeerLICHT & $z$ & 3.276e+14 & 210.88 & 40.79 & 1 \\
0.08293 & MeerLICHT & $q$ & 5.169e+14 & 97.28 & 6.27 & 1 \\
0.08413 & MeerLICHT & $r$ & 4.789e+14 & 114.82 & 14.81 & 1 \\
0.08532 & MeerLICHT & $q$ & 5.169e+14 & 103.76 & 6.69 & 1 \\
0.08659 & MeerLICHT & $i$ & 3.919e+14 & 129.43 & 23.84 & 1 \\
0.08914 & MeerLICHT & $z$ & 3.276e+14 & 307.63 & 62.33 & 1 \\
0.09540 & MeerLICHT & $q$ & 5.169e+14 & 120.23 & 7.75 & 1 \\
0.09662 & MeerLICHT & $r$ & 4.789e+14 & 121.35 & 14.53 & 1 \\
0.10654 & MeerLICHT & $g$ & 6.246e+14 & 60.82 & 9.52 & 1 \\
0.10781 & MeerLICHT & $q$ & 5.169e+14 & 92.90 & 6.85 & 1 \\
0.10899 & MeerLICHT & $r$ & 4.789e+14 & 114.82 & 10.58 & 1 \\
0.11016 & MeerLICHT & $q$ & 5.169e+14 & 93.76 & 6.05 & 1 \\
0.11138 & MeerLICHT & $i$ & 3.919e+14 & 110.67 & 17.33 & 1 \\
0.11252 & MeerLICHT & $q$ & 5.169e+14 & 98.18 & 6.33 & 1 \\
0.11377 & MeerLICHT & $z$ & 3.276e+14 & 157.05 & 40.50 & 1 \\
0.11503 & MeerLICHT & $q$ & 5.169e+14 & 99.09 & 6.39 & 1 \\
0.11754 & MeerLICHT & $q$ & 5.169e+14 & 104.72 & 6.75 & 1 \\
0.11877 & MeerLICHT & $g$ & 6.246e+14 & 69.83 & 10.29 & 1 \\
0.12003 & MeerLICHT & $q$ & 5.169e+14 & 125.90 & 6.96 & 1 \\
0.12121 & MeerLICHT & $r$ & 4.789e+14 & 140.61 & 14.25 & 1 \\
0.12248 & MeerLICHT & $q$ & 5.169e+14 & 108.65 & 10.01 & 1 \\
0.13362 & MeerLICHT & $r$ & 4.789e+14 & 131.83 & 14.57 & 1 \\
0.13482 & MeerLICHT & $q$ & 5.169e+14 & 114.82 & 9.52 & 1 \\
0.13604 & MeerLICHT & $i$ & 3.919e+14 & 134.28 & 25.97 & 1 \\
0.13725 & MeerLICHT & $q$ & 5.169e+14 & 113.77 & 11.53 & 1 \\
0.13958 & MeerLICHT & $q$ & 5.169e+14 & 100.01 & 11.05 & 1 \\
0.14195 & MeerLICHT & $q$ & 5.169e+14 & 127.07 & 8.19 & 1 \\
0.14305 & MeerLICHT & $g$ & 6.246e+14 & 97.28 & 11.65 & 1 \\
0.14430 & MeerLICHT & $q$ & 5.169e+14 & 125.90 & 6.96 & 1 \\
0.14550 & MeerLICHT & $r$ & 4.789e+14 & 159.97 & 14.73 & 1 \\
0.14670 & MeerLICHT & $q$ & 5.169e+14 & 136.78 & 7.56 & 1 \\
0.14797 & MeerLICHT & $i$ & 3.919e+14 & 169.05 & 26.47 & 1 \\
0.15050 & MeerLICHT & $z$ & 3.276e+14 & 275.44 & 48.20 & 1 \\
0.16412 & MeerLICHT & $q$ & 5.169e+14 & 157.05 & 21.70 & 1 \\
0.16650 & MeerLICHT & $q$ & 5.169e+14 & 141.91 & 27.45 & 1 \\
0.16775 & MeerLICHT & $g$ & 6.246e+14 & 119.13 & 24.14 & 1 \\
0.913 & MeerLICHT & $r$ & 4.789e+14 & 25.12 & 4.40 & 1 \\
0.915 & MeerLICHT & $q$ & 5.169e+14 & 24.21 & 1.56 & 1 \\
0.915 & MeerLICHT & $g$ & 6.246e+14 & 16.75 & 3.70 & 1 \\
0.918 & MeerLICHT & $i$ & 3.919e+14 & 41.31 & 8.37 & 1 \\
1.018 & MeerLICHT & $r$ & 4.789e+14 & 21.09 & 4.66 & 1 \\
1.018 & MeerLICHT & $q$ & 5.169e+14 & 23.12 & 1.28 & 1 \\
1.020 & MeerLICHT & $g$ & 6.246e+14 & 17.70 & 2.77 & 1 \\
1.023 & MeerLICHT & $i$ & 3.919e+14 & 32.21 & 8.31 & 1 \\
2.03 & MeerLICHT & $q$ & 5.169e+14 & 7.52 & 0.83 & 1 \\
3.00 & MeerLICHT & $q$ & 5.169e+14 & 3.53 & 0.68 & 1 \\
0.00277 & \Swift{}/UVOT & $white$ & 7.488e+14 & 16.29 & 4.80 & 1 \\
0.13848 & \Swift{}/UVOT & $v$ & 5.511e+14 & 120.23 & 19.93 & 1 \\
0.20686 & \Swift{}/UVOT & $b$ & 6.848e+14 & 92.90 & 7.70 & 1 \\
0.27287 & \Swift{}/UVOT & $uvm2$ & 1.319e+15 & 13.43 & 2.60 & 1 \\
0.340 & \Swift{}/UVOT & $uvw1$ & 1.115e+15 & 16.00 & 2.21 & 1 \\
0.349 & \Swift{}/UVOT & $u$ & 8.583e+14 & 47.43 & 3.93 & 1 \\
0.408 & \Swift{}/UVOT & $uvw2$ & 1.401e+15 & 5.15 & 1.23 & 1 \\
0.418 & \Swift{}/UVOT & $v$ & 5.511e+14 & 87.10 & 12.03 & 1 \\
0.560 & \Swift{}/UVOT & $b$ & 6.848e+14 & 36.31 & 7.69 & 1 \\
0.606 & \Swift{}/UVOT & $uvm2$ & 1.319e+15 & 3.53 & 1.40 & 1 \\
0.650 & \Swift{}/UVOT & $uvw1$ & 1.115e+15 & 5.30 & 1.37 & 1 \\
0.694 & \Swift{}/UVOT & $u$ & 8.583e+14 & 13.80 & 3.94 & 1 \\
0.747 & \Swift{}/UVOT & $uvw2$ & 1.401e+15 & 3.40 & 1.13 & 0 \\
0.757 & \Swift{}/UVOT & $v$ & 5.511e+14 & 27.04 & 11.21 & 1 \\
0.802 & \Swift{}/UVOT & $b$ & 6.848e+14 & 21.09 & 7.19 & 1 \\
2.79 & \Swift{}/UVOT & $uvm2$ & 1.319e+15 & 1.91 & 0.64 & 0 \\
2.90 & \Swift{}/UVOT & $uvw1$ & 1.115e+15 & 2.25 & 0.75 & 0 \\
3.20 & \Swift{}/UVOT & $u$ & 8.583e+14 & 7.24 & 2.41 & 0 \\
3.21 & \Swift{}/UVOT & $b$ & 6.848e+14 & 15.42 & 5.14 & 0 \\
3.21 & \Swift{}/UVOT & $white$ & 7.488e+14 & 3.56 & 1.19 & 0 \\
3.21 & \Swift{}/UVOT & $v$ & 5.511e+14 & 35.32 & 11.77 & 0 \\
3.71 & \Swift{}/UVOT & $uvw2$ & 1.401e+15 & 1.79 & 0.60 & 0 \\
0.18419 & GROND & $g'$ & 6.536e+14 & 116.96 & 1.08 & 1 \\
0.18419 & GROND & $r'$ & 4.820e+14 & 148.60 & 1.37 & 1 \\
0.18419 & GROND & $i'$ & 3.924e+14 & 188.81 & 1.74 & 1 \\
0.18419 & GROND & $z'$ & 3.335e+14 & 227.00 & 2.09 & 1 \\
0.18437 & GROND & $J$ & 2.418e+14 & 307.63 & 5.67 & 1 \\
0.18437 & GROND & $H$ & 1.820e+14 & 428.57 & 7.89 & 1 \\
0.18437 & GROND & $K$ & 1.381e+14 & 554.66 & 15.33 & 1 \\
1.225 & GROND & $g'$ & 6.536e+14 & 13.68 & 0.13 & 1 \\
1.225 & GROND & $r'$ & 4.820e+14 & 18.20 & 0.17 & 1 \\
1.225 & GROND & $i'$ & 3.924e+14 & 22.91 & 0.42 & 1 \\
1.225 & GROND & $z'$ & 3.335e+14 & 27.29 & 0.50 & 1 \\
1.225 & GROND & $J$ & 2.418e+14 & 46.99 & 2.60 & 1 \\
1.225 & GROND & $H$ & 1.820e+14 & 57.02 & 5.78 & 1 \\
1.225 & GROND & $K$ & 1.381e+14 & 58.62 & 17.82 & 1 \\
2.21 & GROND & $g'$ & 6.536e+14 & 5.11 & 0.09 & 1 \\
2.21 & GROND & $r'$ & 4.820e+14 & 6.25 & 0.17 & 1 \\
2.21 & GROND & $i'$ & 3.924e+14 & 7.24 & 0.40 & 1 \\
2.21 & GROND & $z'$ & 3.335e+14 & 9.64 & 0.53 & 1 \\
2.21 & GROND & $J$ & 2.418e+14 & 20.14 & 2.78 & 1 \\
2.21 & GROND & $H$ & 1.820e+14 & 22.29 & 5.75 & 1 \\
2.21 & GROND & $K$ & 1.381e+14 & 59.71 & 19.90 & 0 \\
5.25 & GROND & $g'$ & 6.536e+14 & 1.51 & 0.14 & 1 \\
5.25 & GROND & $r'$ & 4.820e+14 & 1.45 & 0.17 & 1 \\
5.25 & GROND & $i'$ & 3.924e+14 & 1.14 & 0.38 & 0 \\
5.25 & GROND & $z'$ & 3.335e+14 & 1.57 & 0.52 & 0 \\
5.25 & GROND & $J$ & 2.418e+14 & 8.32 & 2.77 & 0 \\
5.25 & GROND & $H$ & 1.820e+14 & 15.56 & 5.19 & 0 \\
5.25 & GROND & $K$ & 1.381e+14 & 49.21 & 16.40 & 0 \\
285.0 & GROND & $g'$ & 6.536e+14 & 0.40 & 0.07 & 1 \\
285.0 & GROND & $r'$ & 4.820e+14 & 0.48 & 0.09 & 1 \\
285.0 & GROND & $i'$ & 3.924e+14 & 0.76 & 0.25 & 0 \\
285.0 & GROND & $J$ & 2.418e+14 & 6.31 & 2.10 & 0 \\
285.0 & GROND & $H$ & 1.820e+14 & 10.97 & 3.66 & 0 \\
285.0 & GROND & $K$ & 1.381e+14 & 27.54 & 9.18 & 0 \\
0.385 & KAIT & $clear$ & 4.722e+14 & 104.72 & 4.82 & 1 \\
1.183 & VLT & $r$ & 4.830e+14 & 18.54 & 0.34 & 1 \\
1.186 & VLT & $g$ & 6.394e+14 & 13.18 & 0.24 & 1 \\
1.189 & VLT & $z$ & 3.124e+14 & 28.84 & 0.80 & 1 \\
2.09 & NOT & $r$ & 4.830e+14 & 6.79 & 0.25 & 1 \\
2.10 & NOT & $z$ & 3.124e+14 & 12.94 & 1.07 & 1 \\
6.08 & NOT & $r$ & 4.830e+14 & 1.05 & 0.18 & 1 \\
\hline
10.8 & MeerKAT & L & 1.4e+09 & 42.0 & 14.0 & 0 \\
18.2 & VLA & C & 6.0e+09 & 136.0 & 32.0 & 1 \\
18.2 & VLA & X & 1.0e+10 & 182.0 & 8.0 & 1 \\
34.1 & MeerKAT & L & 1.4e+09 & 41.1 & 13.7 & 0 \\
34.2 & VLA & C & 6.0e+09 & 140.0 & 10.0 & 1 \\
34.2 & VLA & X & 1.0e+10 & 199.0 & 9.0 & 1 \\
59.7 & MeerKAT & L & 1.4e+09 & 40.2 & 13.4 & 0 \\
67.1 & VLA & C & 6.0e+09 & 102.0 & 9.0 & 1 \\
67.1 & VLA & X & 1.0e+10 & 50.0 & 15.0 & 1 \\
118.0 & VLA & C & 6.0e+09 & 66.0 & 11.0 & 1 \\
118.0 & VLA & X & 1.0e+10 & 37.0 & 7.0 & 1 \\
\end{longtable}
\tablefoot{All times are relative to the \Swift{}/BAT trigger time. X-ray, optical and radio data are separated by horizontal lines. Detections are all at least at the 3$\sigma$ level except for the first UVOT/$white$ detection, which was at the 2.3$\sigma$ level. Optical and radio upper limits are at the 3$\sigma$ level.}

\end{appendix}

\end{document}